\documentclass[12pt,leqno]{amsart}
\usepackage{amssymb}
\usepackage{tikz-cd}
\usepackage{mathrsfs}
\newtheorem{theorem}{Theorem}[section]
\newtheorem{lemma}[theorem]{Lemma}
\newtheorem{corollary}[theorem]{Corollary}

\newtheorem{definition}[theorem]{Definition}
\newtheorem{remark}[theorem]{Remark}

\newtheorem{conjecture}[theorem]{Conjecture}

\newcommand{\edm}[1]{{\color{blue}#1}}
\usepackage{cancel}

\usepackage{ulem} %\sout{text to cancel}

\begin{document}
%%%%%%%%%%%%%%%%%%%%%%%%%%%%%%%%%%%%%%%%%%%%%%%%%%%%%%%%%%
\title{Perturbation Theory and the Quantum Rabi-model}
%%%%%%%%%%%%%%%%%%%%%%%%%%%%%%%%%%%%%%%%%%%%%%%%%%%%%%%%%%
%%%%%%%%%%%%%%%%%%%%%%%%%%%%%%%%%%%%%%%%%%%%%%%%%%%%%%%%%%
\author{Marcello Malagutti, Alberto Parmeggiani*}

\address{Marcello Malagutti
\newline \indent University College London, Department of Mathematics
\newline \indent 25 Gordon St., London, WC1H 0AY, United Kingdom
\newline \indent \textit{Current address:}
\newline \indent University of Warwick--Warwick
\newline \indent Mathematics Institute (WMI)}
\email {Marcello.Malagutti@warwick.ac.uk}

\address{Alberto Parmeggiani
\newline \indent Dipartimento di Matematica, Universit\`a di Bologna
\newline \indent Piazza di Porta San Donato 5, 40126, Bologna, Italy}
\email{alberto.parmeggiani@unibo.it}

\thanks{The authors are members of the Research Group GNAMPA of INdAM}
\thanks{The first author acknowledges the support by  EPSRC Early Career Fellowship: EP/V001760/1.}
\thanks{The second author was partially supported by the Italian Ministry of University and Research, under PRIN2022 (Scorrimento) “Anomalies
in partial differential equations and applications”, 2022HCLAZ8\_002, J53C24002560006.}
\dedicatory{To Masato Wakayama on his 70th birthday}
\date{}

\begin{abstract}
In the first part of the paper we study a perturbative model of the Rabi system of Quantum Optics. We are therefore able to describe, through Rellich's theory,
an analytic expansion of finite families of eigenvalues, of arbitrary fixed length. In particular, we prove that for finite families of eigenvalues the Braak conjecture holds.
In the second part we study the asymptotics of the Weyl spectral counting function of a class of systems that generalize the Quantum Rabi Model
to an $N$-level atom ($N\geq3$) with $n=N-1$ cavity modes of the electromagnetic field, with particular interest in the case $N=3$ and $n=2$.
\end{abstract}
\maketitle
\pagestyle{myheadings}
%%%%%
\markboth{M. MALAGUTTI and A. PARMEGGIANI}{}

\section{Introduction}\label{sec1}
\renewcommand{\theequation}{\thesection.\arabic{equation}}
\setcounter{equation}{0}
The Quantum Rabi Model (QRM), one of the most important models in Quantum Optics, describes the interaction of a 2-level atom and one cavity-mode
electromagnetic field even when the field is not near resonance with the atomic transition and the coupling strength is not weak. 
In fact, it can be seen as the model leading to the Jaynes-Cummings model by rotating wave approximation, which is valid if the field is near resonance
with the atomic transition, and the coupling strength is weak.  (See Braak \cite{Br} and Rabi \cite{Ra1,Ra2}.)

QRM is written as (with $\boldsymbol{a}$ the annihilation operator, $\boldsymbol{a}^\dagger$ the creation operator and with $I_2$ denoting
the $2\times 2$ identity matrix)
\begin{equation}
Q_{\mathrm{Rabi}}=\boldsymbol{a}^\dagger\boldsymbol{a} I_2+g(\boldsymbol{a}^\dagger+\boldsymbol{a})
\left[\begin{array}{cc}0&1\\1&0\end{array}\right]+\Delta\left[\begin{array}{cc}1&0\\0&-1\end{array}\right],
\label{eqRabiSystem}\end{equation}
where $2\Delta$ is the level splitting (i.e. the difference in the energy of the atomic levels), which we choose positive, and $g>0$ describes the strength of the interaction between the photons and the atom.

Being a system on $\mathbb{R}^n$ with polynomial coefficients, based on the Quantum Harmonic Oscillator, it fits in the framework of 
Non-Commutative Harmonic Oscillators (NCHO) introduced by Wakayama and Parmeggiani in \cite{PW1,PW2} and as later extended in Parmeggiani \cite{P-STNCHO, P-MJ},
and more recently in Malagutti and Parmeggiani \cite{MaPa} (and \cite{MaPa1}). However, a difficulty in the geometric analysis of the Rabi system is
that it is not a semiregular globally elliptic system in the sense of \cite{MaPa}, in that, although the symbol is principally scalar, the symbol of the first order part 
(the semiprincipal symbol) cannot be smoothly diagonalized in the phase space
$\mathbb{R}^n\times\mathbb{R}^n$ (as instead is the case for the Jaynes-Cummings system).

The paper is divided into two parts, both of which are perturbative in nature. In the first part we study a perturbation of the classical $1$-d QRM, which can
be  dealt with by the Rellich perturbation theory as in \cite{R} (we chose that approach in place of Kato's one \cite{K} because more direct). In this part
we destroy the homogeneity of the system, which allows us to bring the system into an isometric system to which we may apply perturbation theory. Since all the 
eigenvalues of the unperturbed system (a scalar quantum harmonic oscillator acting on $L^2(\mathbb{R};\mathbb{C}^2)$) are degenerate, we apply Rellich's
approach by assuming some first-order (in the perturbation parameter) nondegeneracy,
described by the way the interaction parameter is approximated by zeros of Laguerre polynomials. That allows us
to control finite segments of the spectrum of the system, for a sufficiently small perturbation parameter, prove that finite segments of the spectrum are given
by \textit{simple} eigenvalues and that in this context Braak's Conjecture for $Q_{\mathrm{Rabi},\varepsilon}$,
a particular case of our system $Q_R$
(see Section \ref{sec4}) holds, at least for arbitrarily long but finite parts of the spectrum. In addition, Rellich's method gives control on the corresponding eigenfunctions. 
We also deal with "degenerate" (in the sense of Rellich) situations, but
in that case we use the simpler method of quasimodes (see Lazutkin \cite{L}) that, however, gives up control on the eigenfunctions but retains control on the spectrum, proving
its simplicity everywhere under suitable assumptions on the third-order effective Hamiltonian.

At any rate, as seen at the end of Section \ref{sec5} we are also able to deal with the general degenerate case, regardless of the
information on the quadratic form appearing at order $2$ in the perturbation expansion of the eigenvalues. Hence, to sum up
the first part of the paper, we prove Braak's conjecture for finite segments of the spectrum, in the sense that 
\textit{for any fixed window $(0,N)$, $N\in\mathbb{N}$, Braak's conjecture holds true in $(0,N)$ for the atomic energy gap $2\Delta$ small.}

\vspace{.3cm}

\noindent More formally, we have the following theorem (proved in Section 6).

\vspace{.3cm}
\noindent\textbf{Theorem.}
	\textit{
	For all $k_0\in \mathbb{N}$ there exists $\varepsilon_0>0$ such that for all $N\leq k_0$ and all $0<\varepsilon\leq \varepsilon_0
	$,  Braak's conjecture (see Section \ref{sec4} and Conjecture \ref{con:BraakConj}) holds true for the shifted perturbed Rabi Hamiltonian
	$\widehat Q_{R,\varepsilon}$ (see \eqref{eq:shiftedRabi}) in all intervals $[N+1/2,N+3/2)$. Moreover, }
	$$\#(\mathrm{Spec}\,\widehat Q_{R,\varepsilon}\cap [N+1/2,N+3/2))\in \{1,2,3\}.$$
In a sense, our result 
may be thought of as a complement of Rudnick's result \cite{Ru}. For generalizations of the Braak Conjecture to asymmetric Quantum Rabi Models,
see the recent paper by Braak, Nguyen, Reyes-Bustos and Wakayama \cite{BNRW}.

In the second part of the paper we describe in the first place some possible generalizations of the QRM to higher dimensions for an interaction with atoms having several  energy levels, where the space dimension $n$ is related to the number $N$ of atomic levels by the equation $n=N-1.$ Then, in the second place, we give
a general theorem about two-term asymptotics of the Weyl function for a quite general class of systems modelled after the QRM. In order to do that, 
we have to cope with the problem of the non-diagonalizability in the symbol class of the first order part of the system. We bypass such a problem by using a perturbation method that brings the perturbed symbol into the SMGES class, for which we may exploit a pseudodifferential diagonalization of the system and the results of \cite{MaPa}. Variational considerations then allow us to get the seeked two-term asymptotics of the Weyl function. In the third place, finally, we show how to apply the latter result to the generalizations of the QRM in dimension 2, that are models of interactions of two photons and a three-level atom.
The result \textit{was so not far known in the two dimensional and three energy levels}.

We end this introduction by giving the plan of the paper.

In Section \ref{secRabi} we describe the Quantum Rabi Model from a physical perspective, highlighting its connection
 with the Jaynes-Cummings model by Rotating Wave Approximation and its importance as a foundational model in Quantum Optics.
In Section \ref{sec2} we recall our setup, and review the global pseudodifferential calculus used especially in the final Section \ref{sec6}.
In Section \ref{sec3} we give the perturbative result, under a non-degeneracy condition, 
on finite initial segments of the spectrum of the QRM. In Section \ref{sec4} we recall Braak's
conjecture and show its validity in the nondegenerate case. We next deal with the degenerate case in Section \ref{sec5}, and show the validity of Braak's conjecture also in this case (hence in all cases). In addition, by the use of quasi-modes, we will construct instances in which we are able to check Braak's conjecture validity in the degenerate case.
In the final Section \ref{sec6} we introduce the generalizations of the QRM and give a perturbative proof of a two-term asymptotics of the Weyl spectral counting
function.

\vspace{.3cm}

\noindent\textbf{Acknowledgements.} We wish to thank the referees for the thorough reading of the paper and the important observations.

%%%%%%%%%%%%%%
\section{Physics of the Quantum Rabi Model}\label{secRabi}

The Quantum Rabi Model (QRM) describes the fundamental interaction between a two-level quantum system (a qubit, atom, or spin-$\frac{1}{2}$) 
and a single bosonic mode (a quantized harmonic oscillator, typically a cavity or vibrational mode) and it captures the minimal setting in which light-matter
interaction is fully treated quantum mechanically.

\subsection{Hamiltonian and basic structure}

A general form for the Hamiltonian of the Quantum Rabi Model is
\begin{equation}
	H = \hbar \omega\, \boldsymbol{a}^\dagger \boldsymbol{a} + \frac{\hbar \Omega}{2}\sigma_z + \hbar g\,\sigma_x (\boldsymbol{a} + \boldsymbol{a}^\dagger),
\end{equation}
where $\omega$ is the frequency of the bosonic mode, $\Omega$ is the energy splitting of the two-level system, $g$ is the coupling strength,
$\boldsymbol{a}, \boldsymbol{a}^\dagger$ are the bosonic annihilation and creation operators, and $\sigma_x, \sigma_z$ are Pauli matrices acting on the two-level system, that is
$$\sigma_x=\left[\begin{array}{cc}
	0 & 1\\
	1 & 0
\end{array}\right],\quad\sigma_y=\left[\begin{array}{cc}
	0 & -i\\
	i & 0
\end{array}\right],\quad\sigma_z=\left[\begin{array}{cc}
	1 & 0\\
	0 & -1
\end{array}\right],$$
Writing $\sigma_\pm=(\sigma_x\pm i\sigma_y)/2$, the interaction term becomes
\begin{equation}
	H_\text{int} = \hbar g(\sigma_+ \boldsymbol{a} +\sigma_- \boldsymbol{a}^\dagger + \sigma_+ \boldsymbol{a}^\dagger + \sigma_- \boldsymbol{a}),
\end{equation}
where the first two terms correspond to excitation exchange (rotating terms) and the latter two correspond to simultaneous 
creation or annihilation of excitations (counter-rotating terms). Neglecting the counter-rotating terms gives the Jaynes--Cummings model 
(as obtained by Jaynes and Cummings in \cite{JaynesCummings}, 
see also the book by Scully and Zubairy \cite{ScullyZubairy}) which, unlike the full Rabi Hamiltonian, preserves the total excitation number.

\subsection{Symmetry and integrability}
Even if the QRM does not preserve excitation number, it possesses a discrete $\mathbb{Z}_2$ parity symmetry given by
\begin{equation}
	\Pi = \sigma_z (-1)^{\boldsymbol{a}^\dagger \boldsymbol{a}},
\end{equation}
which commutes with the Hamiltonian. This leads to adecomposition of the Hilbert space  into the two eigenspaces of
$\Pi$ which will be relevant for the following of this paper since it will be used to distinguish the two
parity sectors of the spectrum. Moreover, this  decomposition enters in an essential way
Braak's description of the model and the conjecture recalled in Section~\ref{sec4} (see \cite{B}).

\subsection{Coupling regimes}

The physics of the QRM depends strongly on the ratio of the coupling strength to the bare frequencies.

\subsubsection{Weak coupling: Jaynes-Cummings regime}

For $g \ll \omega, \Omega$, and near resonance $\omega \approx \Omega$, the rapidly oscillating counter-rotating terms 
can be neglected under the rotating-wave approximation (RWA). The Hamiltonian reduces to
\begin{equation}
	H_\text{JC} = \hbar \omega \boldsymbol{a}^\dagger \boldsymbol{a} + \frac{\hbar \Omega}{2}\sigma_z + \hbar g(\sigma_+ \boldsymbol{a} + \sigma_- \boldsymbol{a}^\dagger),
\end{equation}
which preserves the total excitation number, as showed by Jaynes and Cummings in \cite{JaynesCummings}.

\subsubsection{Ultrastrong coupling}

When $g/\omega \approx 0.1$, the counter-rotating terms become relevant and the RWA breaks down. 
This is the ultrastrong coupling regime, experimentally realized in superconducting circuit QED (see the paper by Niemczyk \cite{Niemczyk}). 
The ground state contains virtual photons and the system exhibits phenomena such as the Bloch--Siegert shift.

\subsubsection{Deep strong coupling}

For $g/\omega \gtrsim 1$, the system enters the deep strong coupling regime~\cite{FornDiazReview}. 
In the limit $g/\omega \to \infty$ the interaction term dominates and the eigenstates are highly entangled superpositions of qubit and oscillator states.

\subsection{Physical interpretation}

The QRM describes the dressing of the two-level system by the quantized field. In weak coupling this dressing leads to small energy shifts, 
while in strong coupling it produces hybrid light-matter excitations (polaritons). The counter-rotating terms allow virtual excitations, 
which influence measurable quantities such as spectra and ground-state photon populations (for a more complete description of the topic 
see Niemczyk, Deppe, Huebl et al. \cite{Niemczyk}, and the survey by Forn--D\'{i}az, Garc\'{i}a-Ripoll, Peropadre et al. \cite{FornDiazReview}).

\subsection{Experimental realizations}

Flexible realizations of the QRM exist in superconducting circuit QED, trapped ions, and optomechanical systems,
where parameters can be tuned to explore ultrastrong and deep strong coupling regimes \cite{Niemczyk,FornDiazReview}.

\section{The setup}\label{sec2}
\renewcommand{\theequation}{\thesection.\arabic{equation}}
\setcounter{equation}{0}
In this section we describe our setting. We deal with globally elliptic systems (see e.g. Helffer \cite{H} or Shubin \cite{S}, or Parmeggiani \cite{P-STNCHO}) with
polynomial coefficients. 

We work in the phase-space $T^\ast \mathbb{R}^n=\mathbb{R}^n\times\mathbb{R}^n=\mathbb{R}^{2n}$, writing
 $X=(x,\xi) \in \mathbb{R}^{2n}$ for the points in $\mathbb{R}^{2n}$. We write for short $\dot{\mathbb{R}}^{2n}=\mathbb{R}^{2n}\setminus \lbrace 0 \rbrace$. 
As usual, we put $\langle X\rangle=(1+|X|^2)^{1/2}$, $D_{x_j}=-i\partial_{x_j}$. The $L^2$-inner product in $L^2(\mathbb{R}^n)$ (or $L^2(\mathbb{R}^n;\mathbb{C}^N)$) is denoted
by $(\cdot,\cdot)_0$.

We next recall the Shubin calculus of global operators (in the context of the Weyl-H\"ormander calculus).

\begin{definition}\label{defSymb}
Let $a \in C^\infty(\mathbb{R}^{2n})$ and $m \in \mathbb{R}$. We say that \textit{$a$ is a global symbol of order $m$}, and write $a \in S^m(\mathbb{R}^{n})$, 
if for all $\alpha\in \mathbb{Z}_+^{2n}$  there exists $C_{\alpha}>0$ such that
\begin{equation}
\label{Shubinsymbol}
|\partial_X^{\alpha}a(X)|\leq C_{\alpha}\langle X\rangle^{m-|\alpha|}, \quad X \in \mathbb{R}^{2n}.
\end{equation}
A symbol $a\in S^m(\mathbb{R}^n)$ ($m\geq 0$) is elliptic when there is $c_0>0$ such that
$$|a(X)|\geq c_0\langle X\rangle^m,\,\,\,\text{ for $X$ large}.$$
\end{definition}

In the H\"ormander-Weyl calculus (see H\"ormander \cite{H3}), $S(\langle\cdot\rangle^m;g)=S^m(\mathbb{R}^n)$ where the H\"ormander metric is given by
$$X\longmapsto g_X=\frac{|dX|^2}{\langle X\rangle^2}.$$

We denote
$$S^{\infty}(\mathbb{R}^{n})=\bigcup_{m \in \mathbb{R}} S^m(\mathbb{R}^{n})$$
and
$$S^{-\infty}(\mathbb{R}^n)=\bigcap_{m \in \mathbb{R}}S^m(\mathbb{R}^{n})=\mathscr{S}(\mathbb{R}^{2n}).$$

If $a \in S^m(\mathbb{R}^n)$, we associate with it a \textit{pseudodifferential operator} using the so-called \textit{Weyl-quantization}.

\begin{definition}
Let $a \in S^m(\mathbb{R}^n)$. The \textit{Weyl-quantized pseudodifferential operator} associated with $a$ is defined as
$$a^\mathrm{w}(x,D)u(x)= (2\pi)^{-n}\iint_{\mathbb{R}^{2n}}e^{i\langle x-y, \xi \rangle} a((x+y)/2,\xi)u(y)dyd\xi, \quad u \in \mathscr{S}(\mathbb{R}^n).$$
\end{definition}

One has that $a^\mathrm{w}(x,D)$  is a linear operator that acts continuously $\mathscr{S}(\mathbb{R}^n)\to\mathscr{S}(\mathbb{R}^n)$ and extends by duality
to a linear continuous operator $\mathscr{S}'(\mathbb{R}^n)\to\mathscr{S}'(\mathbb{R}^n)$.

\begin{definition}
We define the set of global pseudodifferential operators of order $m$ as
$$\Psi^m(\mathbb{R}^n):=\Bigl\lbrace A\colon\mathscr{S}'(\mathbb{R}^n) \xrightarrow[\textit{\tiny cont.}]{\textit{\tiny linear}} \mathscr{S}'(\mathbb{R}^n); 
\ \exists \, a \in S^m(\mathbb{R}^n), \; A=a^\mathrm{w}(x,D) \Bigr\rbrace.$$
We write 
$$\Psi^\infty(\mathbb{R}^n)=\bigcup_{m\in\mathbb{R}}\Psi^m(\mathbb{R}^n),\quad
\Psi^{-\infty}(\mathbb{R}^n)=\bigcap_{m\in\mathbb{R}}\Psi^m(\mathbb{R}^n).$$
We call the elements of  $\Psi^{-\infty}(\mathbb{R}^n)$ smoothing operators: they are those operators in $\Psi^{\infty}(\mathbb{R}^n)$ that have smoothing, i.e.
$\mathscr{S}(\mathbb{R}^n\times\mathbb{R}^n)$, Schwartz kernel.
\end{definition}

The formal adjoint of $a^\mathrm{w}(x,D)$ is simply ${\bar{a}}^\mathrm{w}(x,D)$ and the composition law is written as follows: \textit{If $a\in S^m(\mathbb{R}^n)$ and
$b\in S^{m'}(\mathbb{R}^n)$ then $a^\mathrm{w}(x,D)b^\mathrm{w}(x,D)=(a\sharp b)^\mathrm{w}(x,D)$ where
$a\sharp b\in S^{m+m'}(\mathbb{R}^n)$ is written, for any given $N\in\mathbb{Z}_+$, as
$$a\sharp b(X)=\sum_{k=0}^N\frac{1}{k!}\left(\frac{i}{2}\sigma(D_X;D_Y)\right)^ka(X)b(Y)\bigl|_{X=Y}+r_{N+1}(X),$$
with $r_{N+1}\in S^{m+m'-(N+1)}(\mathbb{R}^n)$ and where
$$\sigma(D_X;D_Y)=\sigma(D_x,D_\xi;D_y,D_\eta)=\langle D_\xi,D_y\rangle -\langle D_\eta,D_x\rangle.$$}

We recall also the important class $S^m_\mathrm{sreg}(\mathbb{R}^n)$ of \textit{semiregular} symbols.

\begin{definition}
We say that $a\in S^m(\mathbb{R}^n)$ belongs to the class $S^m_\mathrm{sreg}(\mathbb{R}^n)$ of semiregular symbols, if
$a$ possesses an asymptotic expansion 
$$a\sim\sum_{j\geq 0}a_{m-j},$$
where each $a_{m-j}$ is homogeneous in $X\in\mathbb{R}^n\times\mathbb{R}^n\setminus\{(0,0)\}$ of degree $m-j,$ that is
$$a_{m-j}(tX)=t^{m-j}a_{m-j}(X),\quad\forall t>0,\,\,\,\forall X\not=0.$$
Asymptotic means that, chosen an excision function $\chi$ (i.e. $\chi\in C^\infty(\mathbb{R}^{2n}),$ $0\leq\chi\leq 1,$ $\chi\equiv 0$ when,
say, $|X|\leq 1/2,$ $\chi\equiv 1$ when, say, $|X|\geq 1$), for every $N\in\mathbb{Z}_+$ we have
$$a-\chi\sum_{j=0}^Na_{m-j}\in S^{m-(N+1)}(\mathbb{R}^n).$$
The terms $a_m,$ $a_{m-1},$ $a_{m-2}$ and $a_{m-3}$ are, respectively, called the principal, semiprincipal, subprincipal, semisubprincipal symbols
of the operator $a^\mathrm{w}(x,D).$
\end{definition}

A fundamental  feature of the H\"ormander-Weyl calculus is its invariance under the action of the affine symplectic group (see H\"ormander \cite{H3}). Given
$\chi\colon\mathbb{R}^n\times\mathbb{R}^n\longrightarrow\mathbb{R}^n\times\mathbb{R}^n$ an affine symplectomorphism, there exists a metaplectic
operator $U_\chi$, unitary in $L^2$ and automorphisms of both $\mathscr{S}$ and $\mathscr{S}'$ 
(uniquely determined modulo a unit complex number) such that
\begin{equation}
U_\chi^* a^\mathrm{w}(x,D)U_\chi=(a\circ\chi)^\mathrm{w}(x,D).
\label{eqSymplecticInvariance}\end{equation}

It is important to recall that semiregular symbols are preserved by the action of the \textit{linear} symplectic group (as
follows from (\ref{eqSymplecticInvariance})).

We will need to consider vector-valued $\psi$dos. The class of symbols we consider is given by
$$S^m(\mathbb{R}^n;\mathsf{M}_N)=S^m(\mathbb{R}^n)\otimes\mathsf{M}_N,$$
where $\mathsf{M}_N$ denotes the space of the $N\times N$ complex matrices. The composition formulas are then to be understood
as compositions rows-times-columns, and the adjoint of a symbol will be given by the adjoint matrix of symbolic entries. Ellipticity is defined by requiring
that $\det a\in S^{mN}(\mathbb{R}^n)$ is elliptic. Analogously, one
has the class of matrix-valued semiregular symbols $S^m_\mathrm{sreg}(\mathbb{R}^n;\mathsf{M}_N)$. In this case, ellipticity is a condition determined only 
by the principal symbol.

It is convenient to recall the definition of the class of SMGES semiregular systems as introduced in \cite{MaPa} (Def. 2.4).

\begin{definition}\label{defSMGES}
A semiregular symbol $a=a^*\in S^m_\mathrm{sreg}(\mathbb{R}^n;\mathsf{M}_N)$ is said to be a semiregular metric globally elliptic system (SMGES)
if it is principally scalar with a globally elliptic principal part,
and whose semiprincipal part is smoothly diagonalizable by unitary elements of $S^0_\mathrm{sreg}(\mathbb{R}^n;\mathsf{M}_N)$
(for $X\not=0$) into \textbf{distinct} blocks that can be ordered in strictly ascending order for all $X\not=0$.
\end{definition}

Next, as regards the $1$-d QRM we have the following setup.
Given $p_2(X)=|X|^2/2$, $X\in\mathbb{R}^2$, we have that $P_0=p_2^\mathrm{w}(x,D)\in\Psi^2(\mathbb{R})$ is the usual quantum harmonic oscillator 
$(D_x^2+x^2)/2,$ $x\in\mathbb{R}.$
We have that $P_0$, realized as an unbounded operator in $L^2(\mathbb{R})$ with maximal domain the Shubin Sobolev space $B^2(\mathbb{R})$ (see \cite{H,S}), 
is self-adjoint and has a discrete spectrum made of eigenvalues with
multiplicity $1$ given by the sequence $\{N+1/2\}_{N\in\mathbb{Z}_+}$, and corresponding eigenfunctions, the ones we will use in our subsequent computations, given by the
Schwartz functions
\begin{equation}
\varphi_N(x)=\left(\frac{x-\partial_x}{\sqrt{2}}\right)^Ne^{-x^2/2}=H_N(x)e^{-x^2/2},\quad N\in\mathbb{Z}_+
\label{eqHermiteFunctions}\end{equation}
(the $H_N$ being the Hermite polynomials of degree $N$),
that form an orthogonal basis of $L^2(\mathbb{R})$ (and of $\mathscr{S}(\mathbb{R})$ and of $\mathscr{S}'(\mathbb{R})$).
One has
$$(\varphi_N,\varphi_{N'})_0=\sqrt{\pi}\,N!\delta_{NN'},\quad N,N'\in\mathbb{Z}_+.$$
Recall also, in passing, an important property of the Hermite polynomials that will be needed shortly: 
\begin{equation}
H_N(x+y)=\sum_{k=0}^N\binom{N}{k}H_k(x)(\sqrt{2}\,y)^{N-k},\quad\forall x,y\in\mathbb{R}.
\label{eqHermiteTaylor}\end{equation}

The Rabi system we consider in this paper is the following. Let $\alpha,\gamma_1,\gamma_2\in\mathbb{R}$ with $\alpha\not=0$ and $\gamma_1>\gamma_2$. 
For $\varepsilon\in[0,1)$ we consider the system ($I_2$ being the $2\times 2$ identity)

\begin{equation}
Q_R=Q_{R,\varepsilon}(x,D)=p_2^\mathrm{w}(x,D)I_2+\alpha x\left[\begin{array}{cc}0&1\\ 1& 0\end{array}\right]+\varepsilon\left[\begin{array}{cc}
\gamma_1& 0\\ 0 & \gamma_2\end{array}\right],
\label{eqRabi}\end{equation}
with maximal domain
$$D_R=\{u\in L^2(\mathbb{R};\mathbb{C}^2);\,\,Q_Ru\in L^2(\mathbb{R};\mathbb{C}^2)\}=B^2(\mathbb{R};\mathbb{C}^2)$$
(hence, it does not depend on $\varepsilon$).
The operator $Q_R$, thought of as an unbounded operator on $L^2$ with domain $D_R$ is then self-adjoint with a compact resolvent. In fact, its symbol
$Q_R(X)$ belongs to $S^m_\mathrm{sreg}(\mathbb{R};\mathsf{M}_2)$ and is globally elliptic, i.e. the principal part of degree $2$ is invertible for $X\not=0$. By the compact embedding
of $D_R$ into $L^2$, the resolvent operator is therefore compact so that $Q_R$ has a real and discrete spectrum made of an increasing sequence of eigenvalues 
with finite multiplicities. 
As we will see, by Rellich's theory, one has that its spectrum can be parametrized by analytic functions of $\varepsilon$ (see Kato
\cite{K}, Thm. 3.9 page 392--393).

Define next the $L^2$-isometries 
$$T_\pm\colon L^2(\mathbb{R})\longrightarrow L^2(\mathbb{R}),\quad u\longmapsto u(\cdot\mp\alpha),$$ 
that are metaplectic operators associated with the symplectomorphisms $\chi_\pm\colon(x,\xi)\longmapsto(x\pm\alpha,\xi)$. Then, since $T_\pm=T_\mp^*$, we have
$$T_\mp p_2^\mathrm{w}(x,D)T_\pm=(D^2+(x\pm\alpha)^2)/2.$$

The following lemma is fundamental.

\begin{lemma}\label{lemmaNormalForm}
There exists a metaplectic operator $U\colon L^2(\mathbb{R};\mathbb{C}^2)\longrightarrow L^2(\mathbb{R};\mathbb{C}^2)$ (which is an automorphism
of $\mathscr{S}(\mathbb{R};\mathbb{C}^2)$ and of $\mathscr{S}'(\mathbb{R};\mathbb{C}^2)$) such that
\begin{equation}
U^*(Q_{R,\varepsilon}+\frac{\alpha^2}{2}I_2)U=A_\varepsilon=A+\varepsilon B=p_2^\mathrm{w}(x,D)I_2+\varepsilon\left[\begin{array}{cc}\beta_1 & \beta_2T_+^2 \\
\beta_2T_-^2 & \beta_1\end{array}\right],
\label{eqRabi1}\end{equation}
where
$$\beta_1=(\gamma_1+\gamma_2)/2,\quad\beta_2=(\gamma_1-\gamma_2)/2.$$
\end{lemma}
\begin{proof}
Let $U_0=\left[\begin{array}{cc} 1&1\\ 1&-1\end{array}\right]/\sqrt{2},$ and $T=\left[\begin{array}{cc}T_+ & 0\\ 0 &T_-\end{array}\right].$ Then
$$U_0^*(Q_{R,\varepsilon}+\frac{\alpha^2}{2}I_2)U_0=\left[\begin{array}{cc}\frac12(D^2+(x+\alpha)^2)&0\\0&\frac12(D^2+(x-\alpha)^2)\end{array}\right]+
\varepsilon\left[\begin{array}{cc}\beta_1&\beta_2\\ \beta_2&\beta_1\end{array}\right]$$
$$=T^*\Bigl(p_2^\mathrm{w}(x,D)I_2+\varepsilon T\left[\begin{array}{cc}\beta_1&\beta_2\\ \beta_2&\beta_1\end{array}\right] T^*\Bigr)T,$$
which proves the lemma with $U=U_0T^*$.
\end{proof}

\begin{remark}
Note that the symbol $t_\pm$ of $T_\pm$ is given by $t_\pm(X)=e^{\mp i\alpha\xi}$ which belongs to $S(1,|dX|^2).$ In particular, for the composition $S(1;|dX|^2)\sharp S^m(\mathbb{R})$
we just have
$$S(1;|dX|^2)\sharp S^m(\mathbb{R})\subset S(\langle\cdot\rangle^m;|dX|^2).$$
Hence there is no asymptotic calculus, but the continuity properties (on $L^2$ and on Shubin's Sobolev spaces, see \cite{H}) keep holding true.
\end{remark}

\subsection{Relation between $Q_{\mathrm{Rabi},\varepsilon}$ and $Q_R$}\label{secRelationRabi}
Strictly speaking, the Rabi system we consider here is a translation by $\frac12 I_2$ of the "genuine" Rabi system (\ref{eqRabiSystem}), that we
may write (replacing $\Delta$ in (\ref{eqRabiSystem}) by  $\varepsilon\Delta$, with $\varepsilon>0$, and putting $g=\alpha/\sqrt{2}$) as
$$Q_{\mathrm{Rabi},\varepsilon}=p_2^\mathrm{w}(x,D)I_2+\alpha x\left[\begin{array}{cc}0&1\\1&0\end{array}\right]+
\varepsilon\Delta\left[\begin{array}{cc}1&0\\0&-1\end{array}\right]-\frac12 I_2,$$
whence
$$Q_{R,\varepsilon}=Q_{\mathrm{Rabi},\varepsilon}+\frac12 I_2,$$
with $\gamma_1=-\gamma_2$ and $\gamma_1=\Delta$.
Therefore up to translating the whole spectrum of $Q_R$ by $1/2$, studying the spectrum of $Q_{R,\varepsilon}$ with 
$\Delta=\gamma_1=-\gamma_2$
is the same as studying the spectrum of $Q_{\mathrm{Rabi},\varepsilon}$.

Using the metaplectic operator $U$ of Lemma \ref{lemmaNormalForm} above, one has that
$$U^*Q_{\mathrm{Rabi},\varepsilon}U=p_2^\mathrm{w}(x,D)I_2+\varepsilon\Delta\left[\begin{array}{cc}0&T_+^2\\ T_-^2&0\end{array}\right]-\frac{1+\alpha^2}{2}I_2.$$

	Since Braak's conjecture deals with the shifted eigenvalues of the Quantum Rabi model, we introduce for convenience the \textbf{shifted perturbed Quantum Rabi Hamiltonian operator}
	\begin{equation} \label{eq:shiftedRabi}
		\widehat Q_{R,\varepsilon}=Q_{R,\varepsilon}+ \frac{\alpha^2}{2}I_2=Q_{\mathrm{Rabi},\varepsilon}+\frac{1+\alpha^2}{2}I_2
		,\,\,\,\varepsilon>0.
	\end{equation}

\section{Perturbation Theory}\label{sec3}
\renewcommand{\theequation}{\thesection.\arabic{equation}}
\setcounter{equation}{0}
By Lemma \ref{lemmaNormalForm}, we will study the spectrum of $A+\varepsilon B.$
To set up Rellich's perturbation theory, we show in the first place that $B$ is $A$-bounded so that we are actually considering a regular family in the Rellich sense.
We have that $D(A)=D(P_0)\otimes\mathbb{C}^2$ and $B=B^*$ is bounded in $L^2.$ Moreover, $A=P_0\otimes I_2$ is an elliptic global $\psi$do, so that, if 
$E\in\Psi^{-2}(\mathbb{R};\mathbb{C}^2)$ is a parametrix for $P_0\otimes I_2$ (that is, $EA=\mathrm{Id}+R$, with
$R\in\Psi^{-\infty}(\mathbb{R})$) the calculus and the continuity properties give that for all $u\in\mathscr{S}(\mathbb{R};\mathbb{C}^2)$ we have
$$|\!|Bu|\!|_0=|\!|B(EA-R)u|\!|_0\leq |\!|BE|\!|_{L^2\to L^2}|\!|Au|\!|_0+|\!|BR|\!|_{L^2\to L^2}|\!|u|\!|_0\leq C_B(|\!|Au|\!|_0+|\!|u|\!|_0),$$
because of the $L^2\to L^2$ continuity of $BE$ and $BR$. Therefore: \textit{The family $\varepsilon\mapsto A+\varepsilon B$ is a holomorphic regular family in the sense of Rellich \cite{R}
(equivalently, of type (A) in the sense of Kato) for all $|\varepsilon|\leq 1/(2C_B)=:\varepsilon_0,$ see \cite{K}. We denote by $I_{\varepsilon_0}$ the corresponding interval centered at $0$.} 
It follows that, by the very nature of $A+\varepsilon B$, it has a compact resolvent for all $\varepsilon$ and hence its eigenvalues 
can be described as analytic functions of $\varepsilon$ (see Kato \cite{K}, 
Thm. 3.9 page 392--393). We may therefore study the analyticity of finite families of eigenvalues of $Q_{R,\varepsilon}$ in $\varepsilon$.

Since $A$ has spectrum $\{N+1/2;\,\,N\in\mathbb{Z}_+\}$ with constant multiplicity $2$, we have to prepare the ground for Rellich's perturbation theory of degenerate
eigenvalues. To that purpose, we have to study the quadratic form defined by $B$ restricted to the eigenspace of $A$ we are studying.

For some $N\in\mathbb{Z}_+$, let $\lambda_0=N+1/2$, $\varphi=\varphi_N$ (see \eqref{eqHermiteFunctions} for the definition of $\varphi_N$), and 
$$E_0=\mathrm{Ker}(A-\lambda_0)=\mathrm{Span}\{\boldsymbol{\phi}_1=\varphi\otimes e_1/|\!|\varphi|\!|_0,\,\,\boldsymbol{\phi}_2=\varphi\otimes e_2/|\!|\varphi|\!|_0\}
\subset\mathscr{S}(\mathbb{R};\mathbb{C}^2),$$ 
where $\{e_1,e_2\}$ is the canonical basis of $\mathbb{C}^2$. If $\pi_{E_0}\colon L^2(\mathbb{R};\mathbb{C}^2)\longrightarrow E_0$ denotes the orthogonal
projection onto $E_0$, we then have a bounded operator $R_0\colon L^2(\mathbb{R};\mathbb{C}^2)\longrightarrow\mathrm{Ran}(A)\subset
L^2(\mathbb{R};\mathbb{C}^2)$ such that
$$\left\{\begin{array}{ll}R_0(A-\lambda_0)=(A-\lambda_0)R_0=1-\pi_{E_0},\\
           R_0\pi_{E_0}=0.\end{array}\right.$$
(Therefore $R_0$ is a parametrix of the globally elliptic operator $A-\lambda_0,$ because $\pi_{E_0}$ is a smoothing term since it has rapidly decreasing
Schwartz kernel.)\\      
Consider the Hermitian sesquilinear form
$(B\cdot,\cdot)_0\bigl|_{E_0\times E_0}$: for linear combinations $u=w_1\boldsymbol{\phi}_1+w_2\boldsymbol{\phi}_2\in E_0$, 
$w=\left[\begin{array}{c}w_1\\ w_2\end{array}\right]\in\mathbb{C}^2$,
and likewise for $u'\in E_0$ with coefficients $w'=\left[\begin{array}{c}w'_1\\ w'_2\end{array}\right]$, we have
$$(Bu,u')_0=\langle Mw,w'\rangle_{\mathbb{C}^2}$$
($\langle\cdot,\cdot\rangle_{\mathbb{C}^2}$ being the Hermitian product in $\mathbb{C}^2$), where
$$M=\left[\begin{array}{cc}\beta_1& \beta_2(T_+\varphi,T_-\varphi)_0/|\!|\varphi|\!|_0^2\\ [8pt]
\beta_2\overline{(T_+\varphi,T_-\varphi)_0}/|\!|\varphi|\!|_0^2 & \beta_1\end{array}\right]=M^*.$$
(Note that $(T_+\varphi,T_-\varphi)_0\in\mathbb{R}$.)
The eigenvalues of $M$ are given by
$$\mu_\pm=\beta_1\pm\beta_2\frac{|(T_+\varphi,T_-\varphi)_0|}{|\!|\varphi|\!|_0^2}.$$
Then $\mu_+=\mu_-=\beta_1=(\gamma_1+\gamma_2)/2$ (which is then independent of $N$) 
iff $\beta_2(T_+\varphi,T_-\varphi)_0=0$, that is, iff $(T_+\varphi,T_-\varphi)_0=0$. So, to have a nondegenerate splitting in the perturbed eigenvalue
we have to study $(T_+\varphi,T_-\varphi)_0$ for a given Hermite eigenfunction $\varphi$ of $A$.

(We decided to keep the modulus of $(T_+\varphi,T_-\varphi)_0$ because then the perturbed eigenvalues $\lambda_{N,\pm}(\varepsilon)$ are ordered, for $\varepsilon$ sufficiently small
depending on $N$.)

\subsection{Study of $(T_+\varphi,T_-\varphi)_0$} We have the following lemma.

\begin{lemma}\label{lemmaHermite}
For all $N,k\in\mathbb{Z}_+$ the following formula holds
\begin{equation}
(T_+\varphi_N,T_-\varphi_k)_0=e^{-\alpha^2}\sum_{j=0}^{\min\{N,k\}}(-1)^{N-j}\binom{N}{j}\binom{k}{j}(\sqrt2\,\alpha)^{N+k-2j}|\!|\varphi_j|\!|_0^2=
\sqrt{\pi}\,e^{-\alpha^2}p_{N,k}(\sqrt2\,\alpha),
\label{eqHermite0}\end{equation}
where
\begin{equation}
p_{N,k}(Z)=\sum_{j=0}^{\min\{N,k\}}(-1)^{N-j}\binom{N}{j}\binom{k}{j}j!Z^{N+k-2j}.
\label{eqLaguerre0}\end{equation}

In particular, for $N=k$,
\begin{equation}
(T_+\varphi_N,T_-\varphi_N)_0=e^{-\alpha^2}\sum_{j=0}^N(-2\alpha^2)^{N-j}\binom{N}{j}^2|\!|\varphi_j|\!|_0^2=:\sqrt{\pi}\,e^{-\alpha^2}p_N(-2\alpha^2),
\label{eqHermite}\end{equation}
where 
\begin{equation}
p_N(Z)=\sum_{j=0}^N\binom{N}{j}^2(N-j)! Z^j=N!\sum_{j=0}^N\binom{N}{j}\frac{Z^j}{j!}.
\label{eqLaguerre}\end{equation}
\end{lemma}
\begin{proof}
In the first place notice that $|\!|\varphi_N|\!|_0^2=\sqrt{\pi}N!.$ Then
$$(T_+\varphi_N,T_-\varphi_k)_0=e^{-\alpha^2}\int e^{-x^2}H_N(x-\alpha)H_k(x+\alpha)dx$$
(by the Taylor expansion of Hermite polynomials (\ref{eqHermiteTaylor}))
$$=e^{-\alpha^2}\sum_{j=0}^N\sum_{j'=0}^k\binom{N}{j}\binom{k}{j'}(-\sqrt2\,\alpha)^{N-j}(\sqrt2\,\alpha)^{k-j'}\int e^{-x^2}H_j(x)H_{j'}(x)dx$$
$$=e^{-\alpha^2}\sum_{j=0}^{\min\{N,k\}}\binom{N}{j}\binom{k}{j}(-1)^{N-j}(\sqrt2\,\alpha)^{N+k-2j}\underbrace{\int H_j(x)^2e^{-x^2}dx}_{=|\!|\varphi_j|\!|_0^2},$$
which concludes the proof.
\end{proof}

The next important observation is that
$$p_N(-Z)=N! L_N(Z),$$
where
$$L_N(Z)=\sum_{j=0}^N(-1)^j\binom{N}{j}\frac{Z^j}{j!}$$
is the $N$-th Laguerre polynomial. Therefore the vanishing of $(T_+\varphi,T_-\varphi)_0$ depends on the zeros of the Laguerre polynomials.
It is known that the zeros of the $L_k$s are dense in $[0,+\infty)$ (see Chihara \cite{C}). Consider
$$\mathsf{Z}:=\{x\in[0,+\infty);\,\,\exists k\in\mathbb{N},\,\,L_k(x)=0\}=\bigcup_{k\in\mathbb{N}}L^{-1}_k(0).$$
Hence $\mathsf{Z}$ is countable and dense in $[0,+\infty)$, whence $[0,+\infty)\setminus Z$ is uncountable. We have that $L_k^{-1}(0)$ is given by
$$0<x_{k,1}<x_{k,2}<\ldots<x_{k,k}.$$
For a given $\delta>0$ let $I_\delta(x_0)=\{x>0;\,\,|x-x_0|<\delta\}.$

\begin{lemma}\label{lemmaRoots}
Let $x_0\in[0,+\infty)\setminus\mathsf{Z}.$ There exists a sequence $\{k_j\}_{j\geq 1}\subset\mathbb{N}$ with $k_j\nearrow+\infty$ as $j\to+\infty$, 
and a sequence $\{\delta_j\}_{j\geq 1}\subset(0,+\infty)$ with $\delta_j\searrow 0$ as $j\to+\infty$,
such that for every $j$, for all $k<k_j$ the polynomials $L_k$ have all the roots at distance at least $\delta_j$ from $x_0.$
\end{lemma}
\begin{proof}
Let $\delta_1=1/10$ and 
$$k_1=\min\{k\in\mathbb{N};\,\,L_k^{-1}(0)\cap I_{\delta_1}(x_0)\not=\emptyset\}.$$
Choose the least $\nu_1$, $1\leq\nu_1\leq k_1$, so that the root $x_{k_1,\nu_1}$ of $L_{k_1}$ is closest to $x_0$ (of course, it cannot be $x_0$ by hypothesis).
Put then
$$\delta_2=\frac{|x_{k_1,\nu_1}-x_0|}{10}<\frac{\delta_1}{10}=10^{-2},$$
and define
$$k_2=\min\{k>k_1;\,\,L_k^{-1}(0)\cap I_{\delta_2}(x_0)\not=\emptyset\}.$$
Choose then $\nu_2,$ $1\leq\nu_2\leq k_2$, such that $x_{k_2,\nu_2}$ is closest to $x_0.$ By induction we therefore have sequences
$\{k_j\}_{j\geq 1},$ $\{\delta_j\}_{j\geq 1},$ and $\{\nu_j\}_{j\geq 1}$ such that $k_j\to+\infty,$ $x_{k_j,\nu_j}\in L_{k_j}^{-1}(0)$ for all $j\geq 1,$
$$\delta_{j+1}=\frac{|x_{k_j,\nu_j}-x_0|}{10}<\frac{\delta_j}{10}<10^{-(j+1)}\searrow 0\,\,\text{\rm as}\,\,j\to+\infty,$$
$$x_{k_j,\nu_j}\xrightarrow[j\to+\infty]{}x_0,\,\,\text{\rm with}\,\,|x_{k_j,\nu_j}-x_0|\searrow 0.$$
This concludes the proof.
\end{proof}

In other words, Lemma \ref{lemmaRoots} makes sure that if $x_0=2\alpha^2\in(0,+\infty)\setminus\mathsf{Z}$ then given any $j\geq 1$ we may find $k_j\geq 1$
such that for all integers $N<k_j$ the Laguerre polynomials $L_N$ do not have zeros at distance less than $\delta_j.$ Hence, by Lemma \ref{lemmaHermite}
for all such $N$ we have that $(T_+\varphi,T_-\varphi)_0\not=0$, whence (as we will see through Rellich's theory)
the perturbations of $\lambda_N(0)=\lambda_0$ (which has multiplicity 2) split into simple eigenvalues $\lambda^{(N)}_{\pm}(\varepsilon)$. Recall, in fact, that
$$\mu_+-\mu_-=2\beta_2\frac{|(T_+\varphi,T_-\varphi)_0|}{|\!|\varphi|\!|_0^2}.$$
Let then $w_\pm=\left[\begin{array}{c}\mathrm{sgn}(T_+\varphi,T_-\varphi)_0\\\pm 1\end{array}\right]/\sqrt{2}$ be the orthonormal 
eigenvectors of $M$ belonging to $\mu_\pm$ and $\boldsymbol{\phi}_\pm=\varphi\otimes w_\pm/|\!|\varphi|\!|_0$ be a corresponding basis of $E_0.$

Using Satz 2, page 365 of Rellich \cite{R} we therefore have proved the following result.

\begin{theorem}\label{thmRellichRabi}
Let $\alpha$ be such that $2\alpha^2\in(0,+\infty)\setminus\mathsf{Z}$, and let $\{k_j\}_{j\geq 1}$ and $\{\delta_j\}_{j\geq 1}$ be the sequences of Lemma \ref{lemmaRoots}.
Fix $k_j.$ Then, there exists $c_j\in(0,1)$ such that for all $N<k_j$ we have analytic functions 
$$(-c_j,c_j)\ni\varepsilon\mapsto\lambda^{(N)}_\pm(\varepsilon)\in\mathbb{R},$$
$$\lambda^{(N)}_\pm(\varepsilon)=\sum_{k=0}^\infty\lambda_{k,\pm}^{(N)}\varepsilon^k,\quad \lambda_{0,\pm}^{(N)}=N+1/2,\,\,\,\lambda_{1,\pm}^{(N)}=\mu_\pm,$$
and 
$$(-c_j,c_j)\ni\varepsilon\mapsto u^{(N)}_\pm(\varepsilon)\in L^2(\mathbb{R};\mathbb{C}^2)$$
$$u^{(N)}_\pm(\varepsilon)=\sum_{k=0}^\infty u_{k,\pm}^{(N)}\varepsilon^k,\quad u^{(N)}_{0,\pm}=\boldsymbol{\phi}_\pm$$
with $u^{(N)}_\pm(\varepsilon)\in D(A+\varepsilon B)$ for all $|\varepsilon|<c_j$ and
$$|\!|u^{(N)}_\pm(\varepsilon)|\!|_0=1,\,\,\forall|\varepsilon|<c_j,\quad (u^{(N)}_{0,\pm},u^{(N)}_{k,\pm})_0\in\mathbb{R},\,\,\forall k\geq 1,$$
and with $u^{(N)}_{k,\pm}\in\mathscr{S}(\mathbb{R};\mathbb{C}^2)$ for all $k$ and convergence of the series in $L^2$,
such that
$$(A+\varepsilon B)u^{(N)}_\pm(\varepsilon)=\lambda^{(N)}_\pm(\varepsilon)u^{(N)}_\pm(\varepsilon),\quad\forall|\varepsilon|<c_j.$$
\end{theorem}

By unitary conjugation we therefore obtain also the following result for the eigenvalues of our Rabi quantum model $Q_R$. 

\begin{corollary}
In particular, for $0\leq N<k_j$ and $0<\varepsilon<c_j$, the first $2k_j$ eigenvalues 
$$\mu_\pm^{(N)}(\varepsilon)=\lambda^{(N)}_\pm(\varepsilon)-\alpha^2/2,\quad 0\leq N<k_j,$$ 
of $Q_{R,\varepsilon}$ are all simple.
\end{corollary}

We will call the $\lambda^{(N)}_{\pm}(\varepsilon)=\mu^{(N)}_{\pm}(\varepsilon)+\alpha^2/2$ the \textit{shifted} eigenvalues of $Q_{R,\varepsilon}$, that is,
the eigenvalues of $\widehat Q_{R,\varepsilon}$.

Recall, at this point, the parity associated with Rabi's system $Q_{\mathrm{Rabi},\varepsilon}$ (but also with our system $Q_{R,\varepsilon}$), is
given by the parity operator
$$\Pi=(-1)^{\boldsymbol{a}^\dagger\boldsymbol{a}}\sigma_z.$$
Since $\Pi$ commutes with $Q_R$ we have an orthogonal splitting 
$$L^2(\mathbb{R};\mathbb{C}^2)=\boldsymbol{H}_+\oplus\boldsymbol{H}_-,$$
where $\boldsymbol{H}_\pm$ are the eigenspaces of the parity operator belonging to the eigenvalues $\pm 1$, respectively.
Recall that the structure of $\boldsymbol{H}_\pm$ is the following:
$$\boldsymbol{H}_+=\left\{\left[\begin{array}{c}f_1\\ f_2\end{array}\right]\in L^2(\mathbb{R};\mathbb{C}^2);\,\,\text{\rm $f_1$ is even and $f_2$ is odd}\right\},$$
$$\boldsymbol{H}_-=\left\{\left[\begin{array}{c}f_1\\ f_2\end{array}\right]\in L^2(\mathbb{R};\mathbb{C}^2);\,\,\text{\rm $f_1$ is odd and $f_2$ is even}\right\}.$$

Now, Theorem \ref{thmRellichRabi} gives also a precise control on the parity of the eigenvalues of $Q_R$ by virtue of the control on the analytic family
of eigenvectors (which are the conjugation of the eigenfunctions $u_\pm^{(N)}(\varepsilon)$ via the metaplectic operator $U$ in Lemma \ref{lemmaNormalForm} ). We start from the case $\varepsilon=0$.

Let $v_{\pm}^{(N)}(\varepsilon):=U_{0}T^{*}u_{\pm}^{(N)}(\varepsilon)$, $0\leq \varepsilon<\varepsilon_0$, 
be the eigenfunctions associated with the eigenvalues $\lambda_{\pm}^{(N)}(\varepsilon)$ of $\widehat Q_{R,\varepsilon} $, respectively, 
which are simple if $\varepsilon\neq 0$. Also, put for short
$$s_N:=\mathrm{sgn}(T_+\varphi,T_-\varphi)_0\quad\text{\rm and}\quad \sigma_N=(-1)^Ns_N.$$

\begin{lemma} \label{l:Parity_0}
In the notation and with the hypotheses of Theorem \ref{thmRellichRabi},
the eigenvalue $\lambda_{+}^{(N)}(0)=\lambda_{-}^{(N)}(0)=N+\frac{1}{2}$
of $\widehat Q_{R,0}$ belongs both to $\mathrm{Spec}\,(\widehat Q_{R,0}|_{\boldsymbol{H}_{+}})$
and to $\mathrm{Spec}\,(\widehat Q_{R,0}|_{\boldsymbol{H}_{-}})$. In fact $v_{\pm}^{(N)}(0)\in\boldsymbol{H}_{\pm\sigma_{N}}$.
\end{lemma}
\begin{proof}
We have that $N+\frac{1}{2}$ has multiplicity $2$ with $v_{\pm}^{(N)}(0)=U_{0}T^{*}u_{\pm}^{(N)}(0)$
generating the eigenspace of $\widehat Q_{R,0}$ associated with $N+\frac{1}{2}$
by Theorem \ref{thmRellichRabi}.

\noindent If, say, $v_+^{(N)}(0)$ had nontrivial projections onto $\boldsymbol{H}_+$ and $\boldsymbol{H}_-$, then also $v_+^{(N)}(\varepsilon)$
would have nontrivial projections onto $\boldsymbol{H}_+$ and $\boldsymbol{H}_-$, for small $\varepsilon,$ whence by orthogonality the multiplicity
of $\lambda_+^{(N)}(\varepsilon)$ would be 2 for $\varepsilon\not=0$ and small, which is impossible. Therefore, $\pm$-respectively, $v_\pm^{(N)}(0)$ belongs to either
$\boldsymbol{H}_+$ or $\boldsymbol{H}_-$.

\noindent Next, a direct computation gives
$$v_{\pm}^{(N)}(0)=\frac{1}{\|\varphi\|_{0}\sqrt{2}}\begin{bmatrix}s_NT_{-}\varphi\pm T_{+}\varphi\\
  s_NT_{-}\varphi\mp T_{+}\varphi\end{bmatrix}$$
and we study when, $\pm$-respectively, $v_{\pm}^{(N)}(0)$ belongs to $\boldsymbol{H}_{+}$, the case $\boldsymbol{H}_{-}$ being analogous.
It is possible if and only if 
$$s_NT_-\varphi\pm T_+\varphi\quad\text{\rm is even and}\,\,
s_NT_-\varphi\mp T_+\varphi\quad\text{\rm is odd.}$$ 
This implies 
$$s_N\varphi(x+\alpha)=\pm\varphi(-x-\alpha),\forall x\in\mathbb{R},$$
which is equivalent to 
$$s_N\varphi(x+\alpha)=\pm(-1)^{N}\varphi(x+\alpha),\forall x\in\mathbb{R},$$
since a Hermite polynomial is even or odd according to whether its degree is even or
odd, respectively. This proves the claim.
\end{proof}

Now we generalize the result to small $\varepsilon>0$.

\begin{lemma} \label{l:Parity_e}
In the notation and with the hypotheses of Theorem \ref{thmRellichRabi}, the eigenvalue
$\lambda_{\pm}^{(N)}(\varepsilon)$ belongs to $\text{\ensuremath{\mathrm{Spec}\,}(\ensuremath{\widehat Q_{R,\varepsilon}|_{\boldsymbol{H}_{\pm\sigma_{N}}}})}\,$
(that is, it has parity $\pm\sigma_{N}$) for all $0<\varepsilon<\varepsilon_{0}$.
\end{lemma}
\begin{proof}
We prove the claim by contradiction. Assume
that $v_{\pm}^{(N)}(\varepsilon')\in\boldsymbol{H}_{\mp\sigma_{N}}$ for
some $0<\varepsilon'<\varepsilon_{0}$. Consider the continuous curves
$[ 0,\varepsilon_{0})\ni\varepsilon\mapsto v_{\pm}^{(N)}(\varepsilon)$
given by Theorem \ref{thmRellichRabi}. Since $v_{\pm}^{(N)}(0)\in\boldsymbol{H}_{\pm\sigma_{N}}$
by Lemma \ref{l:Parity_0}, there is an $\tilde{\varepsilon}$ such
that $v_{\pm}^{(N)}(\tilde{\epsilon})\in\boldsymbol{H}_{+}\cap\boldsymbol{H}_{-}=\{0\}$
by the orthogonality of $\boldsymbol{H}_{+}$ and $\boldsymbol{H}_{-}$.
This is impossible since $v_{\pm}^{(N)}(\tilde{\varepsilon})$ are
eigenvectors.
\end{proof}

\section{The Braak conjecture}\label{sec4}
\renewcommand{\theequation}{\thesection.\arabic{equation}}
\setcounter{equation}{0}
In \cite{B}, Braak conjectured that, within an interval $[N,N+1),$ $N\in\mathbb{Z}_+$, there may be lying either $0,$ or $1$ or at most $2$ shifted eigenvalues $\lambda+g^2=\lambda+\alpha^2/2$,
$\lambda\in\mathrm{Spec}(Q_{\mathrm{Rabi}})$, of the Rabi system. 
Moreover, if in $[N,N+1)$ there are two roots of the function $G_+$ (resp. $G_-$; see \cite{B}) then the intervals $[N-1,N)$ or $[N+1,N+2)$ may contain zero or at most $1$ root of $G_+$
(analogously for $G_-$). Otherwise said (see \cite{Ru}), Braak's $G$-conjecture states the following: \textit{For each parity $N$ determined by the parity operator $\Pi$, 
the interval $[N,N+1)$ contains at most two shifted eigenvalues, two intervals containing no shifted eigenvalues are not adjacent, and two 
intervals containing two shifted eigenvalues are also not adjacent.}

\noindent The latter conjecture has been proved by Rudnick in \cite{Ru} "for almost all $N$s".

\vspace{.3cm}

It is also convenient to restate Braak's conjecture as in Lanuza \cite{Lanuza}: 
\begin{conjecture}[Braak's conjecture]\label{con:BraakConj}
    \textit{Let $\varepsilon>0$ and let
    $B_{N}^{\pm}=\#\:\mathrm{Spec}\,(\widehat Q_{R,\varepsilon}|_{\boldsymbol{H}_{\pm}})\cap\bigl[N+1/2,N+3/2\bigr)$.
    For all $N\geq 0$ and each choice of $\pm$, we have that $B_{N}^{\pm}\in\{0,1,2\}$
    and $B_{N}^{\pm}+B_{N+1}^{\pm}\in\{1,2,3\}$.}
\end{conjecture}

\noindent In addition, as we shall see, we may state: \textit{For $\varepsilon>0$
the interval $\bigl[-1/2,1/2\bigr)$ contains exactly one eigenvalue of parity $-\sigma_0$}.

\vspace{.3cm}

Recall that for the system $Q_{R,\varepsilon}$, we have
by Theorem \ref{thmRellichRabi} that for any given $j\geq 1$ there is a $k_j\in\mathbb{Z}_+$ and $c_j>0$ such that for the first 
$2k_j$ shifted eigenvalues of $Q_{R,\varepsilon}$,
the $\lambda^{(N)}_{\pm}(\varepsilon)$ with $N<k_j,$ we have, for $0\leq\varepsilon<c_j$,
\begin{equation}
\lambda_{N,\pm}(\varepsilon) =N+\frac12+\mu_\pm\varepsilon+o(\varepsilon),\quad N=0,1,\ldots,k_j-1,
\label{eqBraakConj}\end{equation}
or, recalling the definition of $\beta_1$ and $\beta_2$,
$$\lambda_{N,\pm}(\varepsilon)=N+\frac12+\frac12\Bigl(\gamma_1+\gamma_2\pm(\gamma_1-\gamma_2)\frac{|\bigl(T_+\varphi_N,T_-\varphi_N\bigr)_0|}{|\!|\varphi_N|\!|_0^2}\Bigr)\varepsilon
+o(\varepsilon),\quad N=0,\ldots,k_j-1.$$

In view of Theorem \ref{thmRellichRabi}, for $Q_{\mathrm{Rabi},\varepsilon}$ (and small $\varepsilon>0$) we have the following corollary, which proves Braak's Conjecture for
finite segments of the spectrum. Because of the shift by $1/2$ in the spectrum, due to our choice 
of definition of $Q_{R,\varepsilon}$, the intervals in which we have to test Braak's conjecture for the $\lambda_\pm^{(N)}(\varepsilon)$s are given by
$$I_{N+1/2}:=\bigl[N+1/2,N+3/2\bigr),\,\,\,\,N\in\mathbb{Z}_+.$$

\begin{corollary}[Braak's Conjecture]\label{corBraakConj}
If $2\alpha^2\not\in\mathsf{Z}$ and $\gamma_{1}=-\gamma_{2}$, the
Braak conjecture holds for every spectral initial segment of eigenvalues:
For any given $j\geq1$ there exist $k_{j}\in\mathbb{Z}_{+}$ ($k_{j}\nearrow+\infty$)
and $c_{j}>0$ sufficiently small such that the conjecture holds for
the interval $I_{N+1/2}$ for all $0< \varepsilon<c_{j}$ and all $N<k_j-2$. \\
More precisely, when $0<\varepsilon<c_j$ for each interval $\bigl[N+1/2,N+3/2\bigr)$
($0\leq N<k_j-2$) there are exactly two simple eigenvalues of $\widehat Q_{R,\varepsilon}$
with parity $\sigma_{N}$ and $-\sigma_{N+1}$, respectively. In particular, for $0<\varepsilon<c_j$
in $\bigl[0,1/2\bigr)$ there is exactly one simple eigenvalue with parity $-\sigma_{0}$.
\end{corollary}
\begin{proof}
When we are in the QRM case, that is, $\gamma_1=-\gamma_2$ so that $\beta_1=0$ and $\beta_2=\gamma_1$, we have
$\pm\mu_{\pm}>0$ (same choice of $\pm$). Hence, for all $0\leq N<k_j-2$ and for $\varepsilon>0$ sufficiently small,
$\lambda_{+}^{(N)}(\varepsilon),\,\lambda_{-}^{(N+1)}(\varepsilon)\in\bigl[N+1/2,N+3/2\bigr)$
and $\lambda_{+}^{(N+1)}(\varepsilon),\,\lambda_{-}^{(N+2)}(\varepsilon)\in\bigl[N+3/2,N+5/2\bigr)$
and these eigenvalues are simple by Theorem \ref{thmRellichRabi}. Hence, for such $N$s, $B_{N}^{\pm}\leq2$.
Moreover, by Lemma \ref{l:Parity_e}, $\lambda_{+}^{(N)}(\varepsilon)$,
$\lambda_{-}^{(N+1)}(\varepsilon)$, $\lambda_{+}^{(N+1)}(\varepsilon)$, and
$\lambda_{-}^{(N+2)}(\varepsilon)$ have parity $\sigma_{N}$, $-\sigma_{N+1}$,
$\sigma_{N+1}$, and $-\sigma_{N+2}$, respectively. Hence, as
$\sigma_{N+1}$ and $-\sigma_{N+1}$ have different sign, and since there are four eigenvalues in the union
$\bigl[N+1/2,N+3/2\bigr)\cup\bigl[N+3/2,N+5/2\bigr)$
and they are all simple, we proved $1\leq B_N^{\pm}+B_{N+1}^{\pm}\leq3$.

Hence, Braak's conjecture holds.
\end{proof}

Therefore, Braak's conjecture may fail just when we are in a degenerate case, that is when $2\alpha^2$ is a zero of at least a Laguerre polynomial. 
This case will be the topic of next section.

\section{The case $2\alpha^2\in\mathsf{Z}$}\label{sec5}
In this section we tackle the case in which $\alpha$ is such that $2\alpha^2\in\mathsf{Z}$, that is, we have a family (possibly infinite) of Laguerre polynomials $L_k$ that
vanish at $2\alpha^2$. So, we let
$$\mathsf{K}_\alpha:=\{k\in\mathbb{Z}_+;\,\,L_k(2\alpha^2)=0\}.$$
Hence $(T_+\varphi_k,T_-\varphi_k)_0=0$ for all $k\in\mathsf{K}_\alpha$ and Rellich's approach fails in this case.

\begin{remark}\label{remarkKalpha}
  Because of the interlacing property of the zeros of Laguerre's polynomial we have
  $$k\in\mathsf{K}_\alpha\Longrightarrow k-1,k+1\not\in\mathsf{K}_\alpha.$$
\end{remark}

To cope with this case, instead of extending Rellich's theorem to a more degenerate setting,
we prefer to use the techniques of quasi-modes (see Lazutkin \cite{L}), which are sufficiently powerful to give us information on the spectrum,
the drawback being the loss of information on the corresponding eigenfunctions.
Recall that, by Kato's theorem (Thm. 3.9 \cite{K}, pages 393-394) , the eigenvalues, as well as the associated eigenfunctions, are real analytic functions
of $\varepsilon\in I_{\varepsilon_0}$. In this section we shall change a little the notation, by putting $\phi_N=\varphi_N/|\!|\varphi_N|\!|_0$.
Recall that an orthonormal basis for $L^2(\mathbb{R};\mathbb{C}^2)$ is given by $\{\phi_N\otimes w_j\}_{N\in\mathbb{Z}_+,\,\,j=1,2}$ with $\{w_1,w_2\}$ forming an orthonormal basis of $\mathbb{C}^2$.

With $N\in\mathsf{K}_\alpha$ and $\lambda_N=N+1/2$, we look for 

\begin{equation}
\Lambda_{\pm,N}(\varepsilon)=\lambda_N+\varepsilon\mu_++\varepsilon^2\Lambda_{\pm,N}^{(2)}+\varepsilon^3\Lambda_{\pm,N}^{(3)},
\quad u_{\pm,N}(\varepsilon)=\phi_N\otimes w_\pm+\varepsilon u_{\pm,N}^{(1)}+\varepsilon^2 u_{\pm,N}^{(2)}
+\varepsilon^3u_{\pm,N}^{(3)},
\label{eqQuasiModes}\end{equation}
which satisfy 
$$(A+\varepsilon B) u_{\pm,N}(\varepsilon)-\Lambda_{\pm,N}(\varepsilon)u_{\pm,N}(\varepsilon)=O(\varepsilon^4),$$
for $0\leq\varepsilon<\varepsilon_0$. 

\begin{definition}\label{defQuadForm}
  For $N\in\mathsf{K}_\alpha$,
  let $\mu_+=(\gamma_1+\gamma_2)/2$ and let $Q^{(2)}_N,Q^{(3)}_N\colon\mathbb{C}^2\times\mathbb{C}^2\longrightarrow\mathbb{C}$
be the sesquilinear forms
\begin{equation}
Q^{(2)}_N(w,w')=-\Bigl((\mu_+-B)R_0(\mu_+-B)\phi_N\otimes w,\phi_N\otimes w'\Bigr)_0,\quad w,w'\in\mathbb{C}^2.
\label{eqForm}\end{equation}
\begin{equation}
Q^{(3)}_N(w,w')=-\Bigl((\mu_+-B)R_0(\mu_+-B)R_0(\mu_+-B)\phi_N\otimes w,\phi_N\otimes w'\Bigr)_0,\quad w,w'\in\mathbb{C}^2.
\label{eqForm'}\end{equation}\end{definition}

An easy computation gives an explicit form for the matrices associated with $Q^{(2)}_N$ and $Q^{(3)}_N.$ Using the scalar parity operator
$(-1)^{\boldsymbol{a}^\dagger\boldsymbol{a}}$ (that is an isometry) 
and the fact that the reduced resolvent operator $R_0$ is actually a scalar operator
$R_0=r_0\,I_2$ which commutes with the scalar parity operator, one sees that 
$$(T_-^2r_0T_+^2\phi_N,\phi_N)_0=(T_+^2r_0T_-^2\phi_N,\phi_N)_0$$
and
$$(T_-^2r_0T_+^2r_0T_-^2\phi_N,\phi_N)_0=(T_+^2r_0T_\edm{-}^2r_0T_+^2\phi_N,\phi_N)_0,$$
whence it follows that the matrix associated with $Q^{(2)}_N$ and $Q^{(3)}_N$, respectively, are
$$-\beta_\edm{2}^2(T_+^2r_0T_-^2\phi_N,\phi_N)\,I_2,\,\,\text{\rm and}\,\,
-\beta_\edm{2}^3(T_+^2r_0T_-^2r_0T_+^2\phi_N,\phi_N)_0\left[\begin{array}{cc} 0 & 1\\ 1 & 0\end{array}\right].$$

In the next theorem we ``spell out'' a possible choice of the quasi-modes $(\boldsymbol{u}_{\pm,N}(\varepsilon),\Lambda_{\pm,N}(\varepsilon))$, directly suggested by
Rellich's procedure. The proof is just a verification.

\begin{theorem}\label{thmQuasiModes}
Let $N\in\mathsf{K}_\alpha.$ Let $Q^{(3)}_N$ be the Hermitian sesquilinear form (\ref{eqForm'}), that we suppose to be nonzero (hence,
 with eigenvalues $\mu_{\pm,N}^{(3)}=\pm|\beta_2|^3|(T_+^2r_0T_-^2r_0T_+^2\phi_N,\phi_N)_0|$. 
Let $w_\pm\in\mathbb{C}^2$ be corresponding orthonormal vectors belonging to $\mu^{(3)}_{\pm,N}$, respectively. Put in (\ref{eqQuasiModes})
$$\Lambda_{\pm,N}^{(2)}=\Lambda_N^{(2)}=\edm{-}\beta_\edm{2}^2(T_+^2r_0T_-^2\phi_N,\phi_N)_0,\,\,\,
\Lambda_{\pm,N}^{(3)}=\mu_{\pm,N}^{(3)},$$
$$u_{\pm,N}^{(1)}=R_0(\mu_+-B)\phi_N\otimes w_\pm,\,\,u_{\pm,N}^{(2)}=R_0(\mu_+-B)R_0(\mu_+-B)\phi_N\otimes w_\pm,$$
$$u_{\pm,N}^{(3)}=R_0(\mu_+-B)R_0(\mu_+-B)R_0(\mu_+-B)\phi_N\otimes w_\pm+\Lambda_N^{(2)}R_0\edm{^2}(\mu_+-B)\phi_N\otimes w_\pm,$$
and $\boldsymbol{u}_{\pm,N}(\varepsilon):=u_{\pm,N}(\varepsilon)/|\!|u_{\pm,N}(\varepsilon)|\!|_0$.\\
Then 
$$(\boldsymbol{u}_{\pm,N}(\varepsilon),\lambda_{\pm,N}(\varepsilon))_{0\leq \varepsilon<\varepsilon_0}$$ 
is a family of quasi-modes (in the sense of Lazutkin, see \cite{L}) such that 
$$|\!|(A+\varepsilon B)\boldsymbol{u}_{\pm,N}(\varepsilon)-\lambda_{\pm,N}(\varepsilon)\boldsymbol{u}_{\pm,N}(\varepsilon)|\!|_0=F(\varepsilon^4),
\,\,\forall\varepsilon\in[0,\varepsilon_0),$$
where
$$F(\varepsilon^4)=\varepsilon^4\frac{\Bigl|\!\Bigr|(B-\mu_+)u^{(3)}_{\pm,N}-\Lambda_N^{(2)}u^{(2)}_{\pm,N}-
\Lambda_{\pm,N}^{(3)}u_{\pm,N}^{(1)}-\varepsilon\Lambda_N^{(2)}u_{\pm,N}^{(3)}-\varepsilon\Lambda_{\pm,N}^{(3)}u_{\pm,N}^{(2)}
-\varepsilon^2\Lambda_{\pm,N}^{(3)}u_{\pm,N}^{(3)}\Bigl|\!\Bigr|_0}{|\!|u_{\pm,N}(\varepsilon)|\!|_0}.$$
\end{theorem}
Note that
$$|\!|u_{\pm,N}(\varepsilon)|\!|_0=1+e_{\pm,N}(\varepsilon),\,\,\varepsilon\to 0+,$$
where
$$|e_{\pm,N}(\varepsilon)|\leq C\varepsilon,\,\,\,\forall\varepsilon\in[0,\varepsilon_0).$$
As a consequence, we have that there is $C_1>0$ such that
$$|F(\varepsilon^4)|\leq C_1^4\varepsilon^4,\,\,\,\forall\varepsilon\in[0,\varepsilon_0).$$
Since
$$|\Lambda_{+,N}(\varepsilon)-\Lambda_{-,N}(\varepsilon)|=\varepsilon^3|\mu_{+,N}^{(3)}-\mu_{-,N}^{(3)}|=:C_2^3\varepsilon^3,\,\,\,\forall\varepsilon\in[0,\varepsilon_0),$$
choosing $0<\varepsilon<\varepsilon_0$ such that $2C_1^4\varepsilon^4<C_2^3\varepsilon^3,$
we have
$$\Bigl[\Lambda_{+,N}(\varepsilon)-C_1^4\varepsilon^4,\Lambda_{+,N}(\varepsilon)+C_1^4\varepsilon^4\Bigr]\cap
\Bigl[\Lambda_{-,N}(\varepsilon)-C_1^4\varepsilon^4,\Lambda_{-,N}(\varepsilon)+C_1^4\varepsilon^4\Bigr]=\emptyset.$$
By \cite{L} (Prop. 32.1), there must then be an eigenvalue of $A+\varepsilon B$ within each of the $\pm$-intervals
$$\Bigl[\Lambda_{\pm,N}(\varepsilon)-C_1^4\varepsilon^4,\Lambda_{\pm,N}(\varepsilon)+C_1^4\varepsilon^4\Bigr].$$
But since the eigenvalues are analytic in $\varepsilon$ and the multiplicity of the eigenvalue, for $\varepsilon$ small, is at most $2$ (because the projector
$\varepsilon\mapsto -(2\pi i)^{-1}\int_{\Gamma}(\widehat Q_{R,\varepsilon}-z)^{-1}dz$, with $\Gamma=\{|z-(N+1/2)|=\delta\}$ contained in the resolvent set of $\widehat Q_{R,0}$,
which is the projector onto the eigenspace of $N+1/2$ for $\varepsilon=0$, is holomorphic for small $\varepsilon$) we therefore
have that also in this case near $N+1/2$ there are exactly $2$ distinct eigenvalues. By unitary conjugation we therefore have the following theorem.

\begin{theorem} \label{t:DistinctEigen_Degen}
Suppose $2\alpha^2\in\mathsf{Z}$. For each $N\in\mathsf{K}_\alpha$ for which the quadratic form $Q^{(3)}_N$ over $\mathbb{C}^2$ is nonzero, we have that
$\widehat Q_{R,\varepsilon}$, for small $\varepsilon>0$, has exactly $2$ eigenvalues near $N+1/2$, that is, for all small $\delta>0$ there is $\varepsilon_\delta>0$ such that for all $0<\varepsilon<\varepsilon_\delta$,
$$
\#\{\lambda\in \mathrm{Spec}\,(\widehat Q_{R,\varepsilon});\,  |\lambda-(N+1/2)|<\delta\}=2.
$$ 
\end{theorem}

When looking at the spectrum of the QRM near $N+1/2$ for $N\in\mathsf{K}_\alpha$,
we have that $\mu_+=\mu_-=0$ (because $\gamma_1=-\gamma_2$ and $N\in\mathsf{K}_\alpha$),
so that the unperturbed degenerate eigenvalue $\lambda^{(N)}_\pm(0)=N+1/2$ is subjected to an
$\varepsilon^3$ perturbation. When $Q_N^{(3)}$ is nonzero, it necessarily has two eigenvalues with opposite signes, 
and one has again a configuration of the spectrum consistent with Braak's conjecture.

The point is to control the parity of the eigenfunctions near the eigenvalue $N+1/2$ of the unperturbed operator.

\begin{lemma} \label{l:SplitParity_e=00003D0}
For all $N\geq0$ and for any choice of $\pm$, $\Pi_{\boldsymbol{H}_{\pm}}\mathrm{Ker}(\widehat Q_{R,0}-(N+1/2))\neq\{0\}$
where $\Pi_{\boldsymbol{H}_{\pm}}$ is the projection onto $\boldsymbol{H}_{\pm}$.
\end{lemma}
\begin{proof}
In the first place, we know that 
\begin{equation}
\mathrm{Ker}(\widehat Q_{R,0}-(N+1/2))=\mathrm{Span}\,\{U_{0}T^{*}(\phi_{N}\otimes e_{\pm})\}=\mathrm{Span}\,\Big\{(T_{\mp}\phi_{N})\otimes\begin{bmatrix}1\\
\pm1
\end{bmatrix}\Big\},
\label{eq:EigenBasis_e=00003D0}\end{equation}
 where $e_+=\left[\begin{array}{c}1\\0\end{array}\right]$ and $e_{-}=\left[\begin{array}{c}0\\1\end{array}\right]$,
respectively. Next, since $\boldsymbol{H}_{+}\oplus\boldsymbol{H}_{-}=L^{2}$,
it is sufficient to show that $\mathrm{Ker}(\widehat Q_{R,0}-(N+1/2))\not\subset\boldsymbol{H}_{\pm}$
to prove the claim. In fact, by (\ref{eq:EigenBasis_e=00003D0}) having
$\mathrm{Ker}(\widehat Q_{R,0}-(N+1/2))\subset\boldsymbol{H}_{+}$ is equivalent
to $T_{\mp}\phi_{N}$ is even and $\pm T_{\mp}\phi_{N}$
is odd. However, this is impossible since the only function being
both even and odd is $0$. The same argument applies in case of inclusion in $\boldsymbol{H}_{-}$. This concludes the proof.
\end{proof}

The foregoing lemma allows us to obtain a result on the parity of the two distinct eigenvalues in Theorem \ref{t:DistinctEigen_Degen} 
when $\varepsilon>0$, as the next result shows.

\begin{lemma}
In the notation of Theorem \ref{t:DistinctEigen_Degen}, the two distinct eigenvalues
near $N+1/2$ have different parity for small $\varepsilon>0$, that is, either $v^{(N)}_\pm\in\boldsymbol{H}_\pm$
or $v^{(N)}_\pm\in\boldsymbol{H}_\mp$.
\end{lemma}
\begin{proof}
Let $\varepsilon\mapsto v_{\pm}^{(N)}(\varepsilon)$ be the analytic family
of eigenfunctions of $\widehat Q_{R,\varepsilon}$ associated with the
eigenvalues $\lambda_{\pm}^{(N)}(\varepsilon)$ of $\widehat Q_{R,\varepsilon}$. 

As a first step, we show that either $v_{+}^{(N)}(\varepsilon)\in\boldsymbol{H}_{+}$
or $v_{+}^{(N)}(\varepsilon)\in\boldsymbol{H}_{-}$ (we can proceed
similarly for $v_{-}^{(N)}(\varepsilon)$) for all small $\varepsilon>0$.
In fact, $\Pi_{\boldsymbol{H}_{\pm}}v_{+}^{(N)}(\varepsilon)$ also satisfy the eigenvalue equation
with respect to $\lambda_+^{(N)}(\varepsilon),$ since $\widehat Q_{R,\varepsilon}=Q_{\mathrm{Rabi},\varepsilon}+\frac{1+\alpha^2}{2} I_2$ (recall that $\gamma_1=-\gamma_2$)
commutes with the parity operator. If both projections were nonzero, they would be eigenfunctions as well, and
because of their orthogonality they would be linearly independent, 
yielding that $\lambda_+^{(N)}(\varepsilon)$ could not be simple for $\varepsilon>0$.

Next,  as a second step, we show that if $v_{+}^{(N)}(\varepsilon)\in\boldsymbol{H}_{\sigma}$ ($\sigma=1$ or $-1$)
for some small $\varepsilon>0$, then $v_{+}^{(N)}(\varepsilon)\in\boldsymbol{H}_{\sigma}$
for all small $\varepsilon>0$ (we can proceed similarly for $v_{-}^{(N)}$).
Assume, to fix ideas, $v_{+}^{(N)}(\varepsilon')\in\boldsymbol{H}_{+}$.
If there were $\varepsilon'>0$ and $\delta\in (0,\varepsilon')$ 
such that $v_{+}^{(N)}(\varepsilon)\in\boldsymbol{H}_{+}$ for $\varepsilon\in(\varepsilon'-\delta,\varepsilon')$
and $v_{+}^{(N)}(\varepsilon)\in\boldsymbol{H}_{-}$ for $\varepsilon\in(\varepsilon',\varepsilon'+\delta)$,
then by continuity of the family $\varepsilon\mapsto v_{+}^{(N)}(\varepsilon)$ we would have
$0\neq v_{+}^{(N)}(\varepsilon')\in\boldsymbol{H}_{+}\cap\boldsymbol{H}_{-}=\{0\}$
(by the orthogonality of $\boldsymbol{H}_{+}$ and $\boldsymbol{H}_{-}$).

As a final step, we show that the two eigenvalues $v_{\pm}^{(N)}(\varepsilon)$
cannot be in the same space $\boldsymbol{H}_{+}$ or $\boldsymbol{H}_{-}$
for any fixed $\varepsilon$. In fact, if by contradiction we had that both eigenfunctions
$v_{\pm}^{(N)}(\varepsilon)\in\boldsymbol{H}_{+}$
for some $\varepsilon$ (we can proceed similarly
for $\boldsymbol{H}_{-}$), then, by the second step, we would have
$v_{\pm}^{(N)}(0)\in\boldsymbol{H}_{+}$ and, hence, by continuity $\Pi_{\boldsymbol{H}_{-}}\mathrm{Ker}(\widehat Q_{R,0}-(N+1/2))=\Pi_{\boldsymbol{H}_{-}}\mathrm{Span}\,\{v_{\pm}^{(N)}(0)\}=\{0\}$
against Lemma \ref{l:SplitParity_e=00003D0}. This concludes the proof.
\end{proof} 

We may therefore state the following corollary of our approach.

\begin{corollary}[Braak's conjecture, degenerate case]\label{corBraakConj2}
  Consider our system $\widehat Q_{R,\varepsilon}$ in the case of $Q_{\mathrm{Rabi},\varepsilon}$, i.e. when $\mu_\pm=0$. Suppose the Hermitian sesquilinear form $Q_N^{(3)}$ is nonzero.
  Then Braak's conjecture (see Conjecture \ref{con:BraakConj}) keeps holding true for small $\varepsilon>0$.
\end{corollary}

It will be interesting, as suggested by the quasi-mode construction, also to consider cases in which the eigenvalues are either constant (i.e. $N+1/2$ for
some $N\in\mathsf{K}_\alpha$) or have derivatives with respect to $\varepsilon$ at $0$ all zero up to some $k_0\in\mathbb{N}$ (hence $d^{k_0+1}\lambda_\pm^{(N)}/d\varepsilon^{k_0+1}(0)\not=0$; notice that in the QRM case we have $\mu_+=0$).
In that case the behavior will be decided (using, for instance, a quasi-mode construction) when $k_0+1$ is odd, by an analogous Hermitian sesquilinear form similar to $Q_N^{(k_0+1)}$ in the case $k_0=2$.

%BEGIN NEW 
We next show that Braak's conjecture is verified even when we have no information about the eigenvalues of the matrix $Q_{N}^{(3)}$ (at the price of giving up control on
the precision of the location of the eigenvalues). In the first place we show a preliminary interesting result.

\begin{lemma} \label{l:ParityRel}
Let $\zeta_{\pm}=\begin{bmatrix}\zeta_{\pm}^{1}\\ \zeta_{\pm}^{2}\end{bmatrix}\in\mathbb{C}^{2}\setminus\{0\}$ and assume that $\zeta_{\pm}^{1}=\pm(-1)^{N}\zeta_{\pm}^{2}$ for some
$N$. Then $U_{0}T^{*}(\phi_{N}\otimes\zeta_{\pm})$, are eigenvectors of $\widehat Q_{R,0}$ belonging to the eigenvalue $N+1/2$, and
belong to the eigenspace of the parity operator associated with the eigenvalue $\pm1$, $\pm$-respectively.
\end{lemma}
\begin{proof}
It is trivial that $U_{0}T^{*}(\phi_{N}\otimes\zeta_{\pm})$ are both eigenvectors
of $\widehat Q_{R,0}$ associated with the eigenvalue $N+1/2$. We prove the parity property. We have
\[
U_{0}T^{*}(\phi_{N}\otimes\zeta_{\pm})=U_{0}\begin{bmatrix}(T_{-}\phi_{N})\zeta_{\pm}^{1}\\
(T_{+}\phi_{N})\zeta_{\pm}^{2}
\end{bmatrix}=\frac{1}{\sqrt{2}}\begin{bmatrix}(T_{-}\phi_{N})\zeta_{\pm}^{1}+(T_{+}\phi_{N})\zeta_{\pm}^{2}\\
(T_{-}\phi_{N})\zeta_{\pm}^{1}-(T_{+}\phi_{N})\zeta_{\pm}^{2}
\end{bmatrix}.
\]
Hence, by imposing $(T_{-}\phi_{N})\zeta_{+}^{1}+(T_{+}\phi_{N})\zeta_{+}^{2}$ to
be even and $(T_{-}\phi_{N})\zeta_{+}^{1}-(T_{+}\phi_{N})\zeta_{+}^{2}$
to be odd, we have that $U_{0}T^{*}(\phi_{N}\otimes\zeta_{+})$ belongs
to the eigenspace of the parity operator associated with the eigenvalue
$+1$ if and only if $\zeta_{+}^{1}=(-1)^{N}\zeta_{+}^{2}$. Similarly,
to have $U_{0}T^{*}(\phi_{N}\otimes\zeta_{-})$ in the the eigenspace
of the parity operator associated with the eigenvalue $-1$ we need
$(T_{-}\phi_{N})\zeta_{-}^{1}+(T_{+}\phi_{N})\zeta_{-}^{2}$ to be
odd and $(T_{-}\phi_{N})\zeta_{-}^{1}-(T_{+}\phi_{N})\zeta_{-}^{2}$
to be even which is equivalent to $\zeta_{-}^{1}=-(-1)^{N}\zeta_{-}^{2}$,
as claimed.
\end{proof}

Next, as already remarked earlier on, we may arrange the eigenvalues, and eigenfunctions, into analytic families that depend on the parameter $\varepsilon$.
We next know that, near each degenerate eigenvalue $\lambda_\pm^{(N)}(0)=N+1/2=\lambda_0$ we may choose the corresponding eigenfunctions $v_\pm^{(N)}(\varepsilon)$ to belong
to $H_\pm,$ $\pm$-respectively (for $\varepsilon>0$ sufficiently small).

\begin{lemma} \label{l:Eigen_dim-1}
Let $P_{N}(\varepsilon)=-\frac{1}{2\pi i}\int_{\gamma}(\widehat Q_{R,\varepsilon}-z)^{-1}\,dz$ (for a suitable simple curve $\gamma\subset\mathbb{C}$ encircling $N+1/2$) be the spectral projection associated with $\widehat Q_{R,\varepsilon}$ such that $P_N(0)$ is the projection onto the eigenspace of the eigenvalue $N+1/2$. For small $\varepsilon\geq0$, there exist $v_{\pm}^{(N)}(\varepsilon)\in P_N(\varepsilon)L^2(\mathbb{R};\mathbb{C}^2)$
with norm $1$, such that $v_{\pm}^{(N)}(\varepsilon)$ belong to different eigenspaces of the parity operator, and $\varepsilon\mapsto v_{\pm}^{(N)}(\varepsilon)$ is analytic.
\end{lemma}
\begin{proof}
Let $\zeta_{\pm}=\begin{bmatrix}\zeta_{\pm}^{1}\\\zeta_{\pm}^{2}
\end{bmatrix}\in\mathbb{C}^{2}\setminus\{0\}$ with $\zeta_{\pm}^{1}=\pm(-1)^{N}\zeta_{\pm}^{2}$. By Lemma \ref{l:ParityRel},
$U_{0}T^{*}(\phi_{N}\otimes\zeta_{\pm})$ belong to $\boldsymbol{H}_\pm,$ $\pm$-respectively.
Notice that $\varepsilon\mapsto P_{N}(\varepsilon)$ is the restriction to $0\leq\varepsilon<\varepsilon_0$ for some $\varepsilon_0$ of an analytic family defined in a neighborhood of $0$ (see \cite{K}). Define
\[
\tilde{v}_{\pm}^{(N)}(\varepsilon)=P_{N}(\varepsilon)(U_{0}T^{*}(\phi_{N}\otimes\zeta_{\pm})),
\]
which therefore belong to different eigenspaces of the parity operator because $P_{N}(\varepsilon)$ does not change the
parity since $\widehat Q_{R,\varepsilon}$ commutes with the parity operator.
Notice that for small $\varepsilon\geq0$, $\tilde{v}_{\pm}^{(N)}(\varepsilon)\neq0$ since by the continuity of $\varepsilon\mapsto P_{N}(\varepsilon)$
\begin{align*}
|\!|\tilde{v}_{\pm}^{(N)}(\varepsilon)-\tilde{v}_{\pm}^{(N)}(0)|\!|_0 & =|\!|(P_{N}(\varepsilon)-P_{N}(0))U_{0}T^{*}(\phi_{N}\otimes\zeta_{\pm})|\!|_0\\
 & \leq|\!|P_{N}(\varepsilon)-P_{N}(0)|\!|_{L^{2}\to L^{2}}|\!|U_{0}T^{*}(\phi_{N}\otimes\zeta_{\pm})|\!|_0\to0,\text{ as }\varepsilon\to0+.
\end{align*}
Hence $|\!|\tilde{v}_{\pm}^{(N)}(\varepsilon)|\!|_0>0$ for small $\varepsilon\geq0$. Setting $v_{\pm}^{(N)}(\varepsilon)=\tilde{v}_{\pm}^{(N)}(\varepsilon)/|\!|\tilde{v}_{\pm}^{(N)}(\varepsilon)|\!|_0$
proves the claim.
\end{proof}

A direct corollary of Lemma \ref{l:Eigen_dim-1} is the following.

\begin{corollary} \label{c:BraakConjGeneral}
There exists $\sigma\in\{+ 1,-1\}$ such that, for $\varepsilon>0$ sufficiently small,
$$\mathrm{Ker}(\widehat Q_{R,\varepsilon}-\lambda_+^{(N)}(\varepsilon))\cap\boldsymbol{H}_\sigma\neq\{0\}\,\,\,\text{\it and}\,\,\,
\mathrm{Ker}(\widehat Q_{R,\varepsilon}-\lambda_-^{(N)}(\varepsilon))\cap\boldsymbol{H}_{-\sigma}\neq\{0\}.$$
\end{corollary}
\begin{proof}
We prove the claim by contradiction. Assume that the claim is false.
Then, there is small $\varepsilon>0$ such that for all $\sigma\in\{+1,-1\}$,
$\mathrm{Ker}\,(\widehat Q_{R,\varepsilon}-\lambda_{+}^{(N)}(\varepsilon))\cap\boldsymbol{H}_{\sigma}=\{0\}$
or $\mathrm{Ker}\,(\widehat Q_{R,\varepsilon}-\lambda_{-}^{(N)}(\varepsilon))\cap\boldsymbol{H}_{-\sigma}=\{0\}$. Denoting 
$V_{\pm}=\mathrm{Ker}\,(\widehat Q_{R,\varepsilon}-\lambda_{\pm}^{(N)}(\varepsilon))$,
it is equivalent to saying that either $V_{+}\cap\boldsymbol{H}_{+}=\{0\}$
and $V_{-}\cap\boldsymbol{H}_{-}=\{0\}$ or $V_{+}\cap\boldsymbol{H}_{-}=\{0\}$
and $V_{-}\cap\boldsymbol{H}_{+}=\{0\}$, that is,
$$V_{+}\cap\boldsymbol{H}_{+}=\{0\}\text{ and }V_{+}\cap\boldsymbol{H}_{-}=\{0\}$$
or
$$V_{+}\cap\boldsymbol{H}_{+}=\{0\}\text{ and }V_{-}\cap\boldsymbol{H}_{+}=\{0\}$$
or
$$V_{-}\cap\boldsymbol{H}_{-}=\{0\}\text{ and }V_{+}\cap\boldsymbol{H}_{-}=\{0\}$$
or
$$V_{-}\cap\boldsymbol{H}_{-}=\{0\}\text{ and }V_{-}\cap\boldsymbol{H}_{+}=\{0\}.$$
It is impossible that  $V_{+}\cap\boldsymbol{H}_{+}=\{0\}\text{ and }V_{+}\cap\boldsymbol{H}_{-}=\{0\}$
since $\widehat Q_{R,\varepsilon}$ and the parity operator commute. Similarly, it is impossible that $V_{-}\cap\boldsymbol{H}_{+}=\{0\}\text{ and }V_{-}\cap\boldsymbol{H}_{-}=\{0\}$.\\
Thus, we may assume $\mathrm{Ker}\,(\widehat Q_{R,\varepsilon}-\lambda_+^{(N)}(\varepsilon))\cap\boldsymbol{H}_{-}=\{0\}$ and $\mathrm{Ker}\,(\widehat Q_{R,\varepsilon}-\lambda_-^{(N)}(\varepsilon))\cap\boldsymbol{H}_{-}=\{0\}$
(one can proceed in a similar way for the case $\mathrm{Ker}\,(\widehat Q_{R,\varepsilon}-\lambda_+^{(N)}(\varepsilon))\cap\boldsymbol{H}_+=\{0\}$ and $\mathrm{Ker}\,(\widehat Q_{R,\varepsilon}-\lambda_-^{(N)}
(\varepsilon))\cap\boldsymbol{H}_+=\{0\}$) which implies $\mathrm{Ker}\,(\widehat Q_{R,\varepsilon}-\lambda_{\pm}^{(N)}(\varepsilon))\subseteq\boldsymbol{H}_{+}$
since $\widehat Q_{R,\varepsilon}$ and the parity operator commute.
Hence, $P_N(\varepsilon)L^2(\mathbb{R};\mathbb{C}^2)\subseteq\boldsymbol{H}_{+}$
which is against Lemma \ref{l:Eigen_dim-1}.
\end{proof}

Corollary \ref{c:BraakConjGeneral} and the interlacing property of the zeros of the Laguerre polynomials lead to Braak's conjecture verification.

In fact, we consider the two intervals $[N+1/2,N+3/2)$ and $[N+3/2,N+5/2)$.
We study the worst case scenario for the validity of Braak's conjecture,
that is $N,\,N+2\in\mathsf{K}_{\alpha}$. Notice that $N+1\notin\mathsf{K}_{\alpha}$ then,
by the interlacing of the Laguerre polynomials zeros, so that we may exploit Rellich's perturbation theory and have that $\lambda^{(N+1)}_-(\varepsilon)\in[N+1/2,N+3/2)$ and $\lambda^{(N+1)}_+(\varepsilon)\in[N+3/2,N+5/2)$ for $\varepsilon>0$. Suppose hence that for small $\varepsilon>0$,
we have $\lambda_{-}^{(N+1)}(\varepsilon)\in[N+1/2,N+3/2)$ and
may have $\lambda_{\pm}^{(N)}(\varepsilon)\in[N+1/2,N+3/2)$ (which is the worst situation for the validity of Braak's conjecture).
By Corollary \ref{c:BraakConjGeneral}, $\lambda_{\pm}^{(N)}(\varepsilon)$
have different parity, whence we may assume without loss of generality that $\lambda_{\pm}^{(N)}(\varepsilon)$ have parity $\pm 1$, $\pm$-respectively. Assume, just to fix ideas, that $\lambda_{-}^{(N+1)}(\varepsilon)$
has parity $+1$. Then in the interval $[N+1/2,N+3/2)$, $\lambda_+^{(N)}(\varepsilon)$ and $\lambda_{-}^{(N+1)}(\varepsilon)$
have parity $+1$ and $\lambda_-^{(N)}(\varepsilon)$ has parity $-1$. Now, $\lambda_{+}^{(N+1)}(\varepsilon)\in[N+3/2,N+5/2)$
with parity $-1$ by Lemma \ref{l:Parity_e}. Hence, Braak's conjecture
does not hold if and only if $\lambda_{\pm}^{(N+2)}(\varepsilon)\in[N+3/2,N+5/2)$, $\pm$-respectively,
and have the same parity. However, the parity cannot be the same because of
Corollary \ref{c:BraakConjGeneral}.\\
Therefore we established the Braak's conjecture even in the case the Hermitian sesquilinear form $Q^{(3)}_N$ is zero.
\begin{corollary}[Braak's conjecture, general case]
	%\label{corBraakConj2}
	For all $k_0\in \mathbb{N}$ there exists $\varepsilon_0>0$ such that for all $N\leq k_0$ and all $0<\varepsilon\leq \varepsilon_0
	$,  Braak's conjecture (see Section \ref{sec4} and Conjecture \ref{con:BraakConj}) holds true for $\widehat Q_{R,\varepsilon}$ (see \eqref{eq:shiftedRabi}) in all intervals $[N+1/2,N+3/2)$.\\
	Moreover, $\#(\mathrm{Spec}\,\widehat Q_{R,\varepsilon}\cap [N+1/2,N+3/2))\in \{1,2,3\}$.
\end{corollary}

%END NEW

%%%%%%%%%%%%%%%%%%%%%%%%%%%%%%

\section{The Weyl law for some generalizations of the Quantum Rabi Model}\label{sec6}
\renewcommand{\theequation}{\thesection.\arabic{equation}}
\setcounter{equation}{0}

In this section we deal with the asymptotic spectral properties of some generalizations of the Quantum Rabi Model.
We extend to this setting the $2$-term asymptotics of the Weyl spectral counting function obtained in Malagutti and Parmeggiani \cite{MaPa} for semiregular SMGES systems
(see Definition \ref{defSMGES} above) modelled after the Jaynes-Cummings Model. 
In that paper the Jaynes-Cummings Model is written as a semiregular Non-Commutative Harmonic Oscillator, in the SMGES class as introduced in \cite{MaPa}: 
studying the high-energy spectral properties of this class of pseudodifferential systems, a Weyl law and a refined Weyl law (Theorems 7.8 and 7.9 in \cite{MaPa}) 
were obtained. (For the same class, also the meromorphic continuation of the spectral zeta-function \cite{Ma}, and a spectral quasi-clustering property \cite{MaPa1} have been obtained.)
We next show how to obtain a Weyl law for a class of systems that includes and generalizes the QRM.

As already remarked, QRM is not an SMGES, in that, although the system is principally scalar, the semiprincipal symbol cannot be smoothly diagonalized in the class
$S^m(\mathbb{R}^n;\mathsf{M}_N)$.
However, we may construct a perturbation that will be used to find spectral asymptotics of the class of systems we are interested in.

\subsection{\label{subsec:Rabi-model-genralization} Quantum Rabi Model generalizations}
In this section we introduce a class of semiregular NCHOs which generalizes the Quantum Rabi Model. 
For this class we state and prove a Weyl law result in Section \ref{subsec:Non-refined-Weyl-Law} below. 
Let $E_{jk}$ be the $N\times N$ matrix which is $1$ in the $(j,k)$ entry and $0$ otherwise.
In what follows $p_2^\mathrm{w}(x,D)=(|D|^2+|x|^2)/2$ will be the $n$-dimensional quantum harmonic oscillator.
One has the following list of models that generalize the QRM, in the different instances in analogy with the Jaynes-Cummings systems \cite{JaynesCummings}.
%%%%%%

\subsubsection{\label{subsec:Rabi-XiConf} The Quantum Rabi Model for an $N$-level atom
and $n=N-1$ cavity-modes in the \textit{$\Xi$-configuration}}

In this case, for $\alpha_1,\ldots\alpha_{N-1}\in\mathbb{R}\setminus\{0\}$,
$\gamma_{1},\ldots\gamma_{N-1}\in\mathbb{R}$ with $\gamma_{1}\leq\gamma_{2}\leq\ldots\leq\gamma_{N-1}$,
we consider the $N\times N$ system in $\mathbb{R}^{n}$, $n=N-1$,
given by
\[
a^{\mathrm{w}}(x,D)=p_2^{\mathrm{w}}(x,D)I_{N}+\sum_{k=1}^{N-1}\alpha_{k}x_{k}\Bigl(E_{k,k+1}+E_{k+1,k}\Bigr)+\sum_{k=1}^{N-1}\gamma_{k}E_{k+1,k+1}.
\]
In this case, the levels of the atom are given by $0$ and the $\gamma_{k}.$

\subsubsection{\label{subsec:Rabi-LambdaConf}The Quantum Rabi Model for an $N$-level atom
and $n=N-1$ cavity-modes in the \textit{$\bigwedge$-configuration}}

In this case, for $\alpha_{1},\ldots\alpha_{N-1}\in\mathbb{R}\setminus\{0\}$,
$\gamma_{1},\ldots\gamma_{N-1}\in\mathbb{R}$ with $\gamma_{1}\leq\gamma_{2}\leq\ldots\leq\gamma_{N-1}$,
we consider the $N\times N$ system in $\mathbb{R}^{n}$, $n=N-1$,
given by 
\[
a^{\mathrm{w}}(x,D)=p_2^{\mathrm{w}}(x,D)I_{N}+\sum_{k=1}^{N-1}\alpha_{k}x_{k}\Bigl(E_{k,N}+E_{N,k}\Bigr)+\sum_{k=1}^{N-1}\gamma_{k}E_{k+1,k+1}.
\]
In this case, the levels of the atom are given by $0$ and the $\gamma_{k}.$

\subsubsection{\label{subsec:Rabi-NiConf}The Quantum Rabi Model for an $N$-level atom
and $n=N-1$ cavity-modes in the so-called \textit{$\bigvee$-configuration}}

In this case, for $\alpha_{1},\ldots\alpha_{N-1}\in\mathbb{R}\setminus\{0\}$,
$\gamma_{1},\ldots\gamma_{N-1}\in\mathbb{R}$ with $\gamma_{1}\leq\gamma_{2}\leq\ldots\leq\gamma_{N-1}$,
we consider the $N\times N$ system in $\mathbb{R}^{n}$, $n=N-1$,
given by 
\[
a^{\mathrm{w}}(x,D)=p_2^{\mathrm{w}}(x,D)I_{N}+\sum_{k=1}^{N-1}\alpha_{k}x_{k}\Bigl(E_{1,k+1}+E_{k+1,1}\Bigr)+\sum_{k=1}^{N-1}\gamma_{k}E_{k+1,k+1}.
\]
In this case, the levels of the atom are given by $0$ and the $\gamma_{k}.$

\subsection{\label{subsec:Non-refined-Weyl-Law} A Weyl law for the generalized Quantum Rabi Models}
We may state the following Weyl law.

\begin{theorem}\label{thm:Rabi_Weyl_Law}
Let $a=a^{*}\in S^2_\mathrm{sreg}(\mathbb{R}^n;\mathsf{M}_N)$ with
$$a\sim\sum_{j\geq0}a_{2-j},$$
($n=N-1$) where $a_{k}=a_{k}^{*}$ is positively homogeneous of degree $k$. Suppose that $a_{2}=p_2I_N$, where now
$p_2(X)=|X|^2/2,$ $X\in\mathbb{R}^{2n}$, is the $n$-d scalar harmonic oscillator and that on $\dot{\mathbb{R}}^{2n}$ there is a smooth matrix-valued function $b_1$,
positively homogeneous of degree $1$,  such that $a_{\epsilon}:=a+\varepsilon b_1$ is an SMGES for all $\varepsilon\in(0,1)$. Suppose that
$A=a^\mathrm{w}(x,D)$, realized as an unbounded operator in $L^2$ with maximal domain $B^2$, is positive.\\
Then, with $\mathbb{R}\ni\lambda\longmapsto\mathsf{N}_A(\lambda)$ denoting
the spectral Weyl counting function associated with $A$,
$$\mathsf{N}_A(\lambda)=(2\pi)^{-n}\left(N\lambda^{n}\int_{p_2\leq1}dX-
\lambda^{n-1/2}\int_{p_2=1}\mathsf{Tr}(a_{1})\frac{ds}{|\nabla p_2|}\right)+o(\lambda^{n-1/2}),$$
 as $\lambda\rightarrow+\infty$. (Here $ds$ is the Riemannian measure induced on $\{p_2=1\}$ by the $2n$-dimensional Euclidean one.)
\end{theorem}
\begin{proof}
To prove the theorem we use a perturbation argument. We consider $A=a^\mathrm{w}(x,D)$ realized as an unbounded operator in $L^2(\mathbb{R}^n;\mathbb{C}^N)$
with maximal domain $D(A)=B^2(\mathbb{R}^n;\mathbb{C}^N)$, and write $B_1=b_1^\mathrm{w}(x,D)$ realized also as an unbounded operator in $L^2$ with domain
$B^1(\mathbb{R}^n;\mathbb{C}^N)$. We then let $A_\varepsilon=A+\varepsilon B_1$ realized on the domain $D(A_\varepsilon)=D(A)$ (independent of $\varepsilon$).
The point will be to obtain an operator inequality between $A_{\varepsilon}$ and $A$ which,
by the minimax principle, leads to a spectral inequality between $A_{\varepsilon}$ and $A$.
The perturbation $A_\varepsilon$ belongs to the class SMGES for $\varepsilon>0$.
Next, we use Theorem 7.8 (Weyl law) in \cite{MaPa} to have a Weyl
law for $A_{\varepsilon}$. Finally, we can obtain a Weyl law for
$A$ by the spectral inequality just proven. Actually, the Weyl law
for the second order operator $A$ is not refined since the perturbation
has order $1$, whence only the first term after the leading term
of the asymptotics can be determined precisely.

Since $A>0$, we may define the powers $A^{\pm1/4}$ as the unbounded realizations
of globally elliptic $\psi$dos with globally elliptic principal symbols $p_2^{\pm 1/4}I_N\in S^{\pm 1/2}(\mathbb{R}^n;\mathsf{M}_N)$. 
Then for all $\phi\in\mathscr{S}$
\begin{align}
|(A^{-1/4}(A_{\varepsilon}-A)A^{-1/4}\phi,\phi)_{L^{2}}|= & \varepsilon|(A^{-1/4}B_1A^{-1/4}\phi,\phi)_{L^{2}}|\nonumber \\
  \leq & \varepsilon|\!|A^{-1/4}B_1A^{-1/4}|\!|_{L^{2}\rightarrow L^{2}}|\!|\phi|\!|^2_{L^{2}}=\varepsilon C|\!|\phi|\!|^2_{L^{2}},
\label{eq:RabiInequality}\end{align}
where $C:=|\!|A^{-1/4}B_1A^{-1/4}|\!|_{L^{2}\rightarrow L^{2}}$.
Now, by the density of $\mathscr{S}$ in $L^{2}$ and the $L^{2}\rightarrow L^{2}$
boundedness of $A^{-1/4}(A_{\varepsilon}-A)A^{-1/4}$, (\ref{eq:RabiInequality})
holds for all $\phi\in L^{2}$. Note also that $A^{1/4}:D(A^{1/4})\subset L^{2}\rightarrow L^{2}$, $D(A^{1/4})=B^{1/2}(\mathbb{R}^n;\mathbb{C}^N)$,
is an elliptic $\psi$do, $D(A^{1/4})=B^{1/2}(\mathbb{R}^n;\mathbb{C}^N)$. Therefore $A^{1/4}$
is surjective, and we may replace $\phi\in L^{2}$ by $A^{1/4}\psi$ ($\psi\in B^{1/2}$) in (\ref{eq:RabiInequality}), obtaining that for
all $\psi\in B^{2}\subset B^{1/2}$
\begin{equation}
  (A\psi,\psi)_{L^{2}}-\varepsilon C(A^{1/2}\psi,\psi)_{L^{2}}\leq(A_{\varepsilon}\psi,\psi)_{L^{2}}\leq(A\psi,\psi)_{L^{2}}+\varepsilon C(A^{1/2}\psi,\psi)_{L^{2}}.
\label{eq:OperatorRabiInequality}\end{equation}
Let $0<\lambda_{1}\leq\lambda_{2}\leq\ldots$ (respectively, $\lambda_{1,\varepsilon}\leq\lambda_{2,\varepsilon}\leq\ldots$)
be the eigenvalues of $A$ (respectively $A_{\varepsilon}$), repeated
according to multiplicities. By the minimax principle and (\ref{eq:OperatorRabiInequality}) we have
\[
\lambda_{j}-\varepsilon C\sqrt{\lambda_{j}}\leq\lambda_{j,\varepsilon}\leq\lambda_{j}+\varepsilon C\sqrt{\lambda_{j}},\quad\forall j\geq1,
\]
which leads to an estimate for the counting function $\mathsf{N}_{A}$
of $A$ and $\mathsf{N}_{A_{\varepsilon}}$ of $A_{\varepsilon}$.
In fact, for $\varepsilon$ sufficiently small
\begin{align*}
\mathsf{N}_{A_{\varepsilon}}(\lambda):= & \#\{j;\,\,\lambda_{j,\varepsilon}\leq\lambda\}\\
\geq & \#\{j;\,\,\lambda_{j}+\varepsilon C\sqrt{\lambda_{j}}\leq\lambda\}\\
= & \#\{j;\,\,\nu_{\varepsilon,+}^{-1}(\lambda_{j})\leq\lambda\}=\mathsf{N}_{A}(\nu_{\varepsilon,+}(\lambda)),
\end{align*}
 and
\begin{align*}
\mathsf{N}_{A_{\varepsilon}}(\lambda):= & \#\{j;\,\,\lambda_{j,\varepsilon}\leq\lambda\}\\
\leq & \#\{j;\,\,\lambda_{j}-\varepsilon C\sqrt{\lambda_{j}}\leq\lambda\}\\
= & \#\{j;\,\,\nu_{\varepsilon,-}^{-1}(\lambda_{j})\leq\lambda\}=\mathsf{N}_{A}(\nu_{\varepsilon,-}(\lambda)),
\end{align*}
where, with $c_{0}>0$ a lower bound of $\mathrm{Spec}(A)$,
$$\nu_{\varepsilon,\pm}:(c_{0},+\infty)\longrightarrow(\nu_{\varepsilon,\pm}(c_{0}),+\infty),\quad
\lambda\mapsto\frac{\varepsilon^{2}}{2}C^{2}+\lambda\mp\varepsilon C\sqrt{\frac{\varepsilon^{2}}{4}C^{2}+\lambda},$$
is a smooth function which is strictly increasing and invertible for $\varepsilon<\varepsilon_{0}:=2\sqrt{c_0}/C,$ with inverse given by
$$\nu_{\varepsilon,\pm}^{-1}:\lambda\mapsto\lambda\pm\varepsilon C\sqrt{\lambda}.$$
Therefore, for $\varepsilon<\varepsilon_{0}$
\[
\mathsf{N}_{A_{\varepsilon}}\bigl(\nu_{\varepsilon,-}^{-1}(\lambda)\bigr)\leq\mathsf{N}_{A}(\lambda)\leq\mathsf{N}_{A_{\varepsilon}}\bigl(\nu_{\varepsilon,+}^{-1}(\lambda)\bigr),
\]
which is equivalent to
\begin{align}
 & \frac{1}{\lambda^{n-1/2}}\left(\mathsf{N}_{A_{\varepsilon}}\bigl(\nu_{\varepsilon,-}^{-1}(\lambda)\bigr)-(2\pi)^{-n}\lambda^{n}N\int_{p_2\leq1}dX\right)\nonumber \\
\leq & \frac{1}{\lambda^{n-1/2}}\left(\mathsf{N}_{A}(\lambda)-(2\pi)^{-n}\lambda^{n}N\int_{p_2\leq1}dX\right)\nonumber \\
\leq & \frac{1}{\lambda^{n-1/2}}\left(\mathsf{N}_{A_{\varepsilon}}\bigl(\nu_{\varepsilon,+}^{-1}(\lambda)\bigr)-(2\pi)^{-n}\lambda^{n}N\int_{p_2\leq1}dX\right).\label{eq:RabiIneq}
\end{align}
Now, we study the behavior of $\mathsf{N}_{A_{\varepsilon}}\circ\nu_{\varepsilon,+}^{-1}$
when $\lambda\rightarrow+\infty$. By Theorem 7.8 in \cite{MaPa}
and since the semiprincipal symbol of $A_{\varepsilon}$ is $a_{1}+\varepsilon b_1$,
\begin{align*}
\mathsf{N}_{A_{\varepsilon}}\bigl(\nu_{\varepsilon,+}^{-1}(\lambda)\bigr)= & (2\pi)^{-n}\biggl(N\,\nu_{\varepsilon,+}^{-1}(\lambda)^{n}\int_{p_2\leq1}dX\\
 & -\nu_{\varepsilon,+}^{-1}(\lambda)^{n-1/2}\int_{p_2=1}\mathsf{Tr}(a_{1}+\varepsilon b_1)\frac{ds}{|\nabla p_2|}\biggr)+O_\varepsilon(\lambda^{n-1})\\
= & (2\pi)^{-n}\left(\lambda^{n}N\left(\int_{p_2\leq1}dX\right)\right.\\
 & -\lambda^{n-1/2}\biggl(\int_{p_2=1}{\rm \mathsf{Tr}}(a_{1}+\varepsilon b_1)\frac{ds}{|\nabla p_2|}-\varepsilon nCN\int_{p_2\leq1}dX\biggr)\biggr)+O_\varepsilon(\lambda^{n-1}),
\end{align*}
 as $\lambda\rightarrow+\infty$, where the constant in $O_\varepsilon(\lambda)$ may depend on $\varepsilon$. In a similar way, for
$\mathsf{N}_{A_{\varepsilon}}\circ\nu_{\varepsilon,-}^{-1}$ we have
\begin{align*}
\mathsf{N}_{A_{\varepsilon}}(\nu_{\varepsilon,-}^{-1}(\lambda))= & (2\pi)^{-n}\left(\lambda^{n}N\left(\int_{p_2\leq1}dX\right)\right.\\
 & -\lambda^{n-1/2}\biggl(\int_{p_2=1}{\rm \mathsf{Tr}}(a_{1}+\varepsilon b_1)\frac{ds}{|\nabla p_2|}+\varepsilon nCN\int_{p_2\leq1}dX\biggr)\biggr)+O_\varepsilon(\lambda^{n-1}),
\end{align*}
 as $\lambda\rightarrow+\infty$. 

 Hence, since we are looking for the behavior of $\mathsf{N}_{A}$ as $\lambda\rightarrow+\infty$, and since one has
 $O_\varepsilon(\lambda^{n-1})/\lambda^{n-1/2}\xrightarrow[\lambda\to+\infty]{}0$ for all $\varepsilon$, by (\ref{eq:RabiIneq}) we get
\begin{align*}
 & -(2\pi)^{-n}\left(\int_{p_2=1}{\rm \mathsf{Tr}}(a_{1}+\varepsilon b_1)\frac{ds}{|\nabla p_2|}+\varepsilon nCN\int_{p_2\leq1}dX\right)\\
\leq & {\displaystyle \liminf_{\lambda\rightarrow+\infty}}\left(\frac{1}{\lambda^{n-1/2}}\left(\mathsf{N}_{A}(\lambda)-(2\pi)^{-n}\lambda^{n}N\int_{p_2\leq1}dX\right)\right)\\
\leq & {\displaystyle \limsup_{\lambda\rightarrow+\infty}}\left(\frac{1}{\lambda^{n-1/2}}\left(\mathsf{N}_{A}(\lambda)-(2\pi)^{-n}\lambda^{n}N\int_{p_2\leq1}dX\right)\right)\\
\leq & -(2\pi)^{-n}\left(\int_{p_2=1}{\rm \mathsf{Tr}}(a_{1}+\varepsilon b_1)\frac{ds}{|\nabla p_2|}-\varepsilon nCN\int_{p_2\leq1}dX\right),
\end{align*}
 for all $\varepsilon<\varepsilon_{0}$ which implies, by taking
 the limit as $\varepsilon\rightarrow0+$,
 
\begin{align*}
-(2\pi)^{-n}\int_{p_2=1}{\rm \mathsf{Tr}}(a_{1})\frac{ds}{|\nabla p_2|} 
& \leq{\displaystyle \liminf_{\lambda\rightarrow+\infty}}\left(\frac{1}{\lambda^{n-1/2}}\left(\mathsf{N}_{A}(\lambda)-(2\pi)^{-n}\lambda^{n}N\int_{p_2\leq1}dX\right)\right)\\
 & \leq{\displaystyle \limsup_{\lambda\rightarrow+\infty}}\left(\frac{1}{\lambda^{n-1/2}}\left(\mathsf{N}_{A}(\lambda)-(2\pi)^{-n}\lambda^{n}N\int_{p_2\leq1}dX\right)\right)\\
 & \leq-(2\pi)^{-n}\int_{p_2=1}{\rm \mathsf{Tr}}(a_{1})\frac{ds}{|\nabla p_2|},
\end{align*}
 and completes the proof.
\end{proof}
Theorem \ref{thm:Rabi_Weyl_Law} may be used with $N=3$ and $n=N-1=2$ to obtain a non-refined
Weyl law for the models of Subsection \ref{subsec:Rabi-model-genralization}. In fact, taking as $A$ one of the generalized
Rabi models introduced at the beginning of the section (or, more precisely, as its shifting by a constant to make it positive: this is harmless by sharp G\"arding inequality), we show how the term $B$ may be so chosen to have that $A+\varepsilon B$
is indeed an SMGES to which the theorem applies.
Namely, in the notation of Theorem \ref{thm:Rabi_Weyl_Law}:

\begin{itemize}
\item For the models of Subsection \ref{subsec:Rabi-XiConf} we have that 
\[
a=p_2I_{3}+\sum_{k=1}^{2}\alpha_{k}x_{k}\Bigl(E_{k,k+1}+E_{k+1,k}\Bigr)+\sum_{k=1}^{2}\gamma_{k}E_{k+1,k+1},
\]
\[
b_1=\sum_{k=1}^{2}\alpha_{k}\Bigl((i\xi_{k})^{*}E_{k,k+1}+i\xi_{k}E_{k+1,k}\Bigr).
\]
 In fact, 
 \begin{align*}
a+\varepsilon b_1= & p_2I_{3}+\sum_{k=1}^{2}\sqrt{2}\alpha_{k}\Bigl(\psi_{k,\varepsilon}^{*}E_{k,k+1}+\psi_{k,\varepsilon}E_{k+1,k}\Bigr)\\
 & +\sum_{k=1}^{2}\gamma_{k}E_{k+1,k+1},
\end{align*}
 where $\psi_{k,\varepsilon}:=(x_k+i\varepsilon\xi_k)/\sqrt{2},$ $k=1,2,$
is an SMGES for all $\varepsilon>0$;
\item For the models of Subsection \ref{subsec:Rabi-LambdaConf},
Theorem \ref{thm:Rabi_Weyl_Law} can be used to obtain a non-refined Weyl law with
\[
a=p_2I_{3}+\sum_{k=1}^{2}\alpha_{k}x_{k}\Bigl(E_{k,3}+E_{3,k}\Bigr)+\sum_{k=1}^{2}\gamma_{k}E_{k+1,k+1},
\]
\[
b_1=\sum_{k=1}^{2}\alpha_{k}\Bigl((i\xi_{k})^{*}E_{k,3}+i\xi_{k}E_{3,k}\Bigr),
\]
 while for the models of Subsection \ref{subsec:Rabi-NiConf},
Theorem \ref{thm:Rabi_Weyl_Law} can be used with
\[
a=p_2I_{N}+\sum_{k=1}^{2}\alpha_{k}x_{k}\Bigl(E_{1,k+1}+E_{k+1,1}\Bigr)+\sum_{k=1}^{2}\gamma_{k}E_{k+1,k+1},
\]
\[
b_1=\sum_{k=1}^{2}\alpha_{k}\Bigl((i\xi_{k})^{*}E_{1,k+1}+i\xi_{k}E_{k+1,1}\Bigr).
\]
\end{itemize}

\vspace{.5cm}

\noindent\textbf{Data Availability Statement.} All data is provided in full in this paper.

\vspace{.5cm}

\noindent\textbf{Declarations}\\
\textbf{Conflict of interest.} There is no conflict of interest.

%%%%%%%%%%%%%%%%%%%%%%%%%%

\end{document}